\documentclass[a4paper,11pt]{article}
\pdfoutput=1 

\usepackage{jcappub} 

\usepackage[T1]{fontenc} 

\usepackage{graphicx}
\usepackage{natbib}
\usepackage{dcolumn}
\usepackage{caption}
\usepackage{subcaption}
\usepackage{hyperref}
\usepackage{verbatim}
\usepackage{lineno}
\modulolinenumbers[5]
\newcommand{\cosmomc}{\texttt{CosmoMC} }

\newcommand{\cosmoejs}{\texttt{CosmoEJS} }

\newcommand{\be}{\begin{equation}}
\newcommand{\ee}{\end{equation}}
\newcommand{\bea}{\begin{eqnarray}}
\newcommand{\eea}{\end{eqnarray}}

\title{\boldmath Exploring the constraints on cosmological models with \cosmoejs}

\author{J. Moldenhauer,\\}
 \author{F. Cavanna, W. O'toole, and W. Zimmerman}
\affiliation{University of Dallas,\\1845 E. Northgate Dr., Irving, Texas, USA 75062}

\emailAdd{jmoldenhauer@udallas.edu}

\abstract{
We introduce new \cosmoejs modules to improve the investigation of the consequences of constraints on the parameter values of cosmological models.  We use \cosmomc to fit dark energy models and modified gravity models to recent data from the cosmic microwave background measurements of the Planck satellite, baryon acoustic oscillations, supernovae type Ia, Hubble Parameter $H(z)$ measurements, and redshift space distortions.  While the results are in agreement with previous constraints for these models, here, we add an investigation into the dynamics of models with \cosmoejs, an interactive \texttt{Java} package of simulations that allow the user to explore the ramifications of choosing various values for the cosmological parameters of a particular model.  We use the statistical fits attained with \cosmomc to choose the parameters for the models, and then, we visually inspect the plots of the simulated theoretical values for comparisons to the observational values, calculate derived cosmological values, and finally plot the expansion history of cosmological models. These new simulations now include modified gravity cosmological models as well as observations of the growth of structures of galaxies for a more accurate description of the universe's dynamics.  The latest version of \cosmoejs is available from \url{http://www.compadre.org/osp/items/detail.cfm?ID=12406}.
}

\begin{document}

\maketitle
\flushbottom
\section{Introduction \label{Intro}}
Almost two decades after its discovery \cite{SNeDiscovery1,SNeDiscovery2,SNeDiscovery3}, the cause of the cosmic acceleration, or accelerated expansion of the universe on cosmological scales of distances, still poses one of the most intriguing problems facing cosmology.  Choices for explaining the cosmic acceleration range from a cosmological constant, $\Lambda$, or other form of repulsive dark energy, i.e. negative pressure and negative equation of state, to a modification of general relativity at cosmological distances \cite{IshakReview2007}.  While support for the most simple cosmological constant, or $\Lambda$ Cold Dark Matter ($\Lambda$CDM) model comes from comparisons to recent cosmological observations \cite{PLA2015}, more precise measurements point to signs of tension between the ($\Lambda$CDM) model and certain data sets \cite{RiessH0}.

Constraints on the parameter values of the most popular cosmological models are straightforward to achieve these days with recent observational data sets.  Some of the more frequently used fitting programs include \cosmomc\cite{CosmoMCPaper}, \texttt{CosmoNest}\cite{CosmoNest}, \texttt{CosmoPMC} \cite{CosmoPMC} and others.  Where the results from a program like \cosmomc are relatively quick (depending on the model, number of data sets and computer), the majority of the output is statistical rather than visually dynamic.  Some web-based simulations \cite{iCosmo} and mobile device applications \cite{CosmoCalcApp} are available for generating dynamical graphs for comparison to data, but do not include the large diversity of data sets or dark energy and modified gravity models.  \texttt{CosmoFish} \cite{CosmoFish} allows one to create forecasts of future data with Fisher matrices and other technical codes \cite{EFTCAMB,MGCAMB,MGCAMB2} allow for the fitting of several extensions or alternatives to general relativity and dark energy, but again are built more for finding constraints on model parameters and not exploring the model's dynamics.  Textbooks  \cite{Rindler,Schneider,BergstromGoobar,Moore} and cosmology calculators \cite{WrightCalculator} that deal with the expansion of the universe, usually include dynamical graphs of the expansion rate, age, or angular diameter distance, but rarely cover the dynamics of the more general dark energy models or even modified gravity models. 

 \cosmoejs is an interactive package of \texttt{Java} simulations that allow the user to visually and numerically compare theoretical cosmological models to experimental data \cite{CosmoEJSPaper}.  As a numerical comparison to the statistical fitting of programs like \cosmomc, the \cosmoejs package uses different $\chi^2$ statistical fitting depending on the data set, see Sec. \ref{DatainCosmoEJS}.  A newly released expanded version of the \cosmoejs package allows for inspections of the visual and numerical fits to be more consistent with the latest fitting programs, i.e. \cosmomc, by adding new data sets and observables.  This latest version also allows for the first time, the simultaneous testing of modified gravity models and dark energy models with observational data sets and their dynamically evolving expansion history.  A dynamical study of dark energy and modified gravity models simultaneously with actual data allows us to see where the models may have different curves and trajectories than each other at high or low redshift.  This version of the \cosmoejs package can be downloaded from \url{http://www.compadre.org/osp/items/detail.cfm?ID=12406} and still requires little technical expertise to use, compared to the majority of the programs mentioned earlier in this section.  

Outline:

The purpose of this article is to introduce the latest version of \cosmoejs, which now includes modified gravity models and more recent data sets, and present an example of the type of comparative analysis that is possible when combining fitting programs like \cosmomc with \cosmoejs.  We describe the classes of models available for testing by our modified versions of \cosmoejs and \cosmomc in Sec. \ref{models}. In Sec. \ref{DatainCosmoEJS} we outline some of the common observations currently used to constrain the parameter values of theoretical models in cosmology.  We find constraints on some of the more popular dark energy models and modified gravity models in Sec. \ref{UsingCosmoMC}.  We explore these values in Sec. \ref{UsingCosmoEJS} using \cosmoejs, and we discuss the advantages and disadvantages to using a program like \cosmoejs in combination with programs like \cosmomc to further the understanding of how parameters in theoretical models affect the dynamics of the model.  In the dynamical plots of \cosmoejs, we will see how the differences in the trajectories of the cosmological models may affect the fit of the precise of low and high redshifts data points.  Our work in this paper concludes in Sec. \ref{Conclusion}, including plans for future developments of the simulations.


\section{Dark Energy and Dvali, Gabadadze and Porrati (DGP) models \label{models}}
We test two particular classes of cosmological models by constraining free parameters in the models with \cosmomc and show their dynamics in comparison to the data with \cosmoejs.  Einstein's field equations or general relativity describe the relationship between the spacetime curvature and the matter or energy density.  By adding a $\Lambda g_{\alpha \beta}$ term to the equations, general relativity can account for the cosmic acceleration with a cosmological constant, $\Lambda$, as a dark energy or repulsive force,
\be 
G_{\alpha \beta} +\Lambda g_{\alpha \beta} = \frac{8\pi G}{\rho} T_{\alpha \beta}.
\label{eq:GR}
\ee
 More generally, the dark energy can be parameterized by describing it as a cosmic fluid, with an equation of state $w=P/\rho$ that varies in redshift.  Allowing for curvature in the metric, and parameterizing the dark energy, $w(z) = w_0+w_a[z/(1+z)]$, the Friedmann equation describing the expansion rate of the universe is written as \cite{CPL},
 \be
 E(z)^2 \equiv \Big(\frac{H(z)}{H_0}\Big)^2 = \Omega_m(1+z)^3+\Omega_\Lambda\Big[(1+z)^{3(1+w_0+w_a)}\exp\Big(\frac{-3 w_a z}{1+z}\Big)\Big] + \Omega_k (1+z)^2 + \Omega_r (1+z)^4,
 \label{eq:FriedmannEquationH}	
 \ee
 with $\Omega_m = \Omega_c + \Omega_b$.  In this equation, $H_0$ is the Hubble constant, $w_0$ is the equation of state, $w_a$ is its derivative,  $\Omega_x$ represents the matter ($m$), cold dark matter ($c$), baryon ($b$), dark energy ($\Lambda$), curvature ($k$), and radiation ($r$) densities, respectively.  We can construct other models as special cases of eq. (\ref{eq:FriedmannEquationH}), i.e. setting $w_0=-1$ and $w_a=0$, reduces to the case of a cosmological constant, and a flat universe is achieved by fixing $\Omega_k = 0.0$, where 
 \be
\Omega_k=-\frac{k}{a^2H^2},
\label{eq:Curvaturedensity}
\ee
and $k$ is the curvature ($k=-1,0,1$ for open, flat, closed, respectively), for spatial geometry.

In order to illustrate comparative dynamical analysis combined with statistical model fitting, we choose to compare one class of models that has historically matched well to observations, i.e. GR dark energy \cite{PLA2015} to a class of models that not only have an unphysical constraint (ghost mode in self accelerating models \cite{DGPghostmode}), but also have been shown to have a tension with observational data sets \cite{Fang2008}.  This way, we know one of our classes of models (DGP) will have significant difficulties that can be more easily compared.  Due to the extensive popularity of the Dvali, Gabadadze and Porrati (DGP) class of models for comparison studies \cite{IshakDEDGP2006, Dossett2010}, and our motivation to emphasize dynamical differences for models that have different statistical fits, we compare its constraints and dynamics to that of the general relativity dark energy (GRDE) class.  The DGP model is described by a five dimensional action that distinguishes between the five dimensional bulk and the four dimensional brane by defining a characteristic length, $r_c$, for which on scales much smaller than $r_c$, gravity appears four dimensional while the complete five dimensional physics is recovered on scales larger than $r_c$ \cite{DGP}.  An effective energy density is defined as,
 
 \be
 \rho_{r_c} \equiv \frac{3}{(32\pi G r_c^2)},
 \label{eq:DGPdensity}
 \ee
 and for comparison to observations, the Friedmann equation for DGP models is given as,
 \be
 E(z)^2 =  \Big[\sqrt{\Omega_m(1+z)^3 + \Omega_{r_c} + \Omega_r (1+z)^4}  + \sqrt{\Omega_{r_c}} \Big]^2+ \Omega_k (1+z)^2  + \Omega_r (1+z)^4   ,
 \label{eq:FriedmannDGP}
 \ee
where $\Omega_{r_c} \equiv 1/(4r_c^2 H_0^2)$ and $\sqrt{\Omega_{r_c}} = (1-\Omega_m-\Omega_r -\Omega_k)/(2\sqrt{1-\Omega_k})$.
 Again, we build the flat case of the DGP universe by setting $\Omega_k=0.0$. We compare both classes of models to observations.  In Sec. \ref{UsingCosmoMC}, we compare the models using precision statistical likelihoods built with covariance matrices, where we expect significant differences in the fitting for the two classes of models.  In Sec. \ref{UsingCosmoEJS}, we graph the evolution and dynamics of models simultaneously with data and error bars, to visualize the differences in the dynamical behavior of models preferred by the tests in Sec. \ref{UsingCosmoMC} .  
  
\section{Cosmological Observations \label{DatainCosmoEJS}}

Many of the research fitting programs already include a sampling of recent observational data sets or methods to generate new ones.  Here, we include the observational phenomenon that we employ with our modified versions of \cosmomc and \cosmoejs.  Both programs contain modules that allow for comparing models to expansion history data and the growth history of structure formations.  Specifically, we use versions of the programs that contain supernovae type Ia (SNeIa) data sets, Cosmic Microwave Background (CMB) radiation distance priors data sets, baryon acoustic oscillation (BAO) data sets, gamma-ray burst (GRB) data sets, Hubble Parameter, $H(z)$, data sets, the Alcock-Paczynski (AP) test data sets, and the growth rate factor, $f(z)$ or $f(z)\sigma_8 $ data sets from redshift space distortions.    Due to the spectrum of different data sets accessible through \cosmoejs, the package does not have the capability to test models with data using the covariance matrix likelihoods or the full CMB Planck data.  As discussed below, \cosmoejs can use a general minimum chisquare $\chi^2$ for all data sets, except for the CMB distance priors, which uses a likelihood and covariance matrix generated using the full CMB Planck data and other data sets in \cosmomc.  We do use covariance matrix likelihoods when statistically fitting the models to observations in Sec. \ref{UsingCosmoMC}, and we use the mean values of the free parameters in these models to set the initial parameters used to calculate the dynamics in \cosmoejs.  In this way, because we have already found the best-fit values of parameters and $\chi^2$ in \cosmomc using the covariance matrix methods, the dynamics in \cosmoejs represent these best-fit models, and we do not present the general minimum chisquare $\chi^2$ calculation in \cosmoejs as it is less accurate for some observations, i.e. SNeIa.  For the most recent list of data sets available, as well as their descriptions, please visit the corresponding webpage for the most stable versions of either programs used here, \cosmoejs: \url{http://www.compadre.org/osp/items/detail.cfm?ID=12406} \footnote{In the case of \cosmoejs, several observations and datasets were also described in \cite{CosmoEJSPaper} and supplemental documents online.} or \cosmomc: \url{http://cosmologist.info/cosmomc/readme.html}.

\subsection{Supernovae Type Ia}
Since the 1998 discovery of the cosmic acceleration, SNeIa have grown in precision and number, and are consistently used to compliment other cosmic observables in describing the expansion history of the universe by forming a redshift-distance relation, see below.  Recent compilations of data sets of SNeIa in \cosmoejs and \cosmomc, have SNeIa that number in the  hundreds  \cite{Union2.1,JLA}, but future releases have been proposed which could detect thousands of measurements in one survey \cite{LSST}.  In order to compare the theoretical models to the observations of SNeIa, we calculate the extinction-corrected distance modulus, $\mu(z)$,

\be
\mu(z)=5\log_{10}[D_L(z)/{\rm Mpc}]+25
\label{eq:distancemodulus}
\ee
where $z$ is the redshift, and $D_{L}$ is the luminosity distance.  $D_L$ has the usual relation,

\begin{equation}
D_L(z)=\frac{1+z}{H_0\sqrt{|\Omega_{k}|}} {\mathcal{S}}\left[\sqrt{|\Omega_{k}|}\int_0^z
\frac{dz'}{H(z')}\right],
\label{eq:LuminosityDistance}
\end{equation}
where

\be
{\mathcal{S}}[\sqrt{|\Omega_k|}x]=\begin{cases}
\sin(\sqrt{|\Omega_k|}x), {\rm if}\ \Omega_k<0\,(k=+1),\\
x,  {\rm if}\  \Omega_k=0\,(k=0),\\
\sinh(\sqrt{|\Omega_k|}x),  {\rm if}\  \Omega_k>0\,(k=-1).
\end{cases}
\ee

The SNeIa data sets available in \cosmomc have their own likelihood methods (see for example \cite{JLA}), where the distance modulus $\mu(z) = m_{B} - (M_{B}-\alpha\times X_1+\beta\times C)$ depends on $m_B$, the observed peak magnitude, $B$ band, with $\alpha$, $\beta$ and $M_B$ nuisance parameters.  When fitting our models, we use the latest version of \cosmomc containing the full covariance likelihood method described in \cite{JLA} and \cite{PLA2015}, to find parameter mean values for the GRDE and DGP cosmological models.   The main emphasis in \cosmoejs is on the visual and dynamical over a statistical model fitting, but \cosmoejs can calculate the simple $\chi^2_{SNeIa}$,

\begin{equation}
\chi^2_{SNeIa}=\sum_{i=1}^{N}\frac{[\mu_{th}(z_i)-\mu_{obs}(z_i)]^2}{\sigma^2_i},
\label{eq:SNechisquare}
\end{equation}
where $\sigma_i$ is the total uncertainty for each SNeIa measurement.  
This general $\chi^2$ is not used for fitting in \cosmoejs, as it is a more relative comparison and can be applied to a large variety of SNeIa data sets \cite{CosmoEJSPaper} without adjustment.  
\footnote{We acknowledge that this general $\chi^2$ method will unfairly weight more accurate and precise nearby SNeIa over the less numerous distant ones, but the \cosmoejs approach is to study the dynamics of particular models and not just to statistically weight one model over another, see Appendix of \cite{CosmoEJSPaper} for more detail.}  
\subsection{Gamma-ray Bursts}

In addition to SNeIa, \cosmoejs has GRB data sets that also measure the expansion history, but at high redshift ($z > 1.4$).  However, due to extinction effects and a small quantity of photons, the uncertainties in these data sets are significantly larger than SNeIa, so, GRB were not included in the fits calculated from \cosmomc.  Again, we stress that \cosmoejs is used for exploring the dynamics, and it is interesting to consider how the models fit data of expansion history at high redshifts, which do on the average show similar expansion behavior to the overlapping SNeIa.  For evaluation to GRB with the theoretical models, \cosmoejs includes the similar $\chi^2_{GRB}$ with eqs. (\ref{eq:distancemodulus}),(\ref{eq:LuminosityDistance}),(\ref{eq:SNechisquare}), however we acknowledge the error bars of the GRB are large and not reliable for significant fitting at high redshift.
\subsection{Hubble Parameter, $H(z)$}
The \cosmoejs package and our modified \cosmomc package contain measurements of the Hubble Parameter, $H(z)$, compiled from several surveys, as listed in Table \ref{Hzdata}, where the user can choose which survey or combination of surveys to include.  The measurement on $H(z)$ provides an independent check of the expansion history coming from galaxy surveys, either directly using the cosmic chronometers method or derived from BAO measurements \cite{HzCCvBAO}, rather than nearby SNeIa or very distant GRBs.  We include BAO and AP measurements in the next sub sections, so we only use the CC data sets for $H(z)$ measurements in \cosmomc testing to avoid the redundancy in data sets.  Again, to keep it similar for different surveys, we use the comparison, $\chi^2_{H(z)}$ as,
\be
\chi^2_{H(z)}=\sum_{i=1}^{N}\frac{[H(z)_{th}-H(z)_{obs}]^2}{\sigma^2_i}.
\label{eq:chisqHz}
\ee

\begin{table}
\centering
\begin{tabular}{| c | c | c | c || c | c | c | c |}
\hline
Redshift  & H(z) & Error & Ref. & Redshift  & H(z) & Error & Ref.\\ \hline
0.07 & 69.0 & 19.6 & \cite{Zhangs}&0.5929 & 104 & 13 & \cite{MaStro}\\		
0.1 & 69.0 & 12.0 & \cite{Ages}&0.6 & 87.9 & 6.1 & \cite{BlakeHz}\\		
0.12 & 68.6 & 26.2 & \cite{Zhangs}&0.6797 & 92 & 8 & \cite{MaStro}\\		
0.17 & 83 & 8 & \cite{Ages}&0.73 & 97.3 & 7.0 & \cite{BlakeHz}\\			
0.1791 & 75 & 4 & \cite{MaStro}&0.7812 & 105 & 12 & \cite{MaStro}\\		
0.1993 & 75 & 5 & \cite{MaStro}&0.8754 & 125 & 17 & \cite{MaStro}\\		
0.2 & 72.9 & 29.6 & \cite{Zhangs}&0.88 & 90 & 40 & \cite{Ages}\\		
0.24 & 79.69 & 2.32 & \cite{SDSSHz}&0.9 & 117 & 23 & \cite{Ages}\\		
0.27 & 77 & 14 & \cite{Ages}&1.037 & 154 & 20 & \cite{MaStro}\\			
0.28 & 88.8 & 36.6 & \cite{Zhangs}&1.3 & 168 & 17 & \cite{Ages}\\		
0.3 & 81.7 & 5.0 & \cite{OkaHz}&1.363 & 160 & 33.6 & \cite{M2015}\\			
0.35 & 82.7 & 8.4 & \cite{ChuangHzSDSSR7}&1.43 & 177 & 18 & \cite{Ages}\\		
0.3519 & 83 & 14 & \cite{MaStro}&1.53 & 140 & 14 & \cite{Ages}\\		
0.4 & 95 & 17 & \cite{Ages}&1.75 & 202 & 40 & \cite{Ages}\\			
0.43 & 86.45 & 3.27 & \cite{SDSSHz}&1.965 & 186.5 & 50.4 & \cite{M2015}\\		
0.44 & 82.6 & 7.8 & \cite{BlakeHz}&2.3 & 224 & 8 & \cite{BDR}\\		
0.48 & 97 & 60 & \cite{Ages}&2.34 & 222 & 7 & \cite{DelubacHz}\\			
0.57 & 96.8 & 3.4 & \cite{AndersonHzBOSS}&2.36 & 226 & 8 & \cite{FontRiberaBOSSDR11}\\ \hline		
\end{tabular}
\caption{$H(z)$ measurements collected from several galaxy surveys.  In \cosmoejs, specific combinations of the data are available by selecting particular surveys, but here, we list all that are included in our versions of \cosmoejs and \cosmomc.  It should be noted that due to differences in measurement techniques (either BAO or Cosmic Chronometers(CC)), caution should be used when combining $H(z)$ measurements from different data sets \cite{HzCCvBAO} because in contrast to the CC data sets, the $H(z)$ measurements from BAO data sets are computed by assuming a cosmological model.}
\label{Hzdata}
\end{table}
\subsection{Baryon Acoustic Oscillations}

Baryon acoustic oscillations (BAO) provide another method to test the cosmic history by comparing the ratio of the sound horizon at the drag epoch, $r_s(z_d)$, or when the baryons decoupled from the primordial universe, to the effective distance, $D_V(z)$ at a late-time effective redshift in the galaxy redshift surveys.  The inclusion of various generations of BAO data sets requires \cosmoejs to use the constraint equation,
\be
\chi^2_{BAO}=\sum_{i=1}^{N}\frac{[\frac{r_s(z_d)}{D_V(z_i)}_{th}-\frac{r_s(z_d)}{D_V(z_i)}_{obs}]^2}{\sigma^2_i},
\label{eq:chisqbao}
\ee
with different effective redshift data points from different surveys \cite{PercivalBAO,WiggleZBAO,Beutler6dFGSBAO}. However, for our statistical fits in Sec. \ref{UsingCosmoMC}, the latest version of \cosmomc utilizes full covariance likelihoods described in \cite{PLA2015,Percival2007,PercivalBAO,AndersonBAOBOSS,WiggleZBAO,Beutler6dFGSBAO}. Comparatively this baryon decoupling occurs at somewhat lower redshift and later time than the photon decoupling because the baryons are embedded in gravitational potential wells.  The correlations in the galaxy redshift surveys consistently have a `bump' at $\approx 102\,h^{-1}$ Mpc, \cite{Percival2007,PercivalBAO,AndersonBAOBOSS,WiggleZBAO,Beutler6dFGSBAO} corresponding to the standard ruler measurement of the BAO. 

The sound horizon is defined as \cite{Percival2007},
\be
r_s(z_d)=\frac{1}{\sqrt{3}}\int^{1/(1+z_d)}_{0}{\frac{da}{a^2H(a)\sqrt{1+(3\Omega_b/4\Omega_{\gamma})a}}},
\label{eq:SoundHorizon}
\ee
where fractional photon energy density, $\Omega_{\gamma}=2.469\times 10^{-5}h^{-2}$ for a temperature of the CMB as $T_{cmb}=2.725 K$ \cite{PLA2015}.
The drag epoch redshift, $z_d$ is \cite{EisensteinHu1998}
\be
z_d=\frac{1291(\Omega_m h^2)^{0.251}}{1+0.659(\Omega_m h^2)^{0.828}}[1+b_1(\Omega_b h^2)^{b_2}],
\label{eq:zdrag}
\ee
where
\be
b_1=0.313(\Omega_m h^2)^{-0.419}[1+0.607(\Omega_m h^2)^{0.674}],
\label{eq:zdragb1}
\ee
and
\be
b_2=0.238(\Omega_m h^2)^{0.223}.
\label{eq:zdragb2}
\ee
The effective distance, $D_V(z)$, according to \cite{Eisenstein2005} is given as
\be
D_V(z)=\Big(D_A^2(z)(1+z)^2\frac{z}{H(z)}\Big)^{1/3},
\label{eq:EffectiveDistance}
\ee
where $D_A(z)$ is the usual proper time angular diameter distance $D_A(z)=D_L(z)/(1+z)^2$. 

\subsection{Cosmic Microwave Background radiation}

The cosmic microwave background (CMB) radiation from the latest Planck Survey is the most precise cosmological observation, to date, of the fluctuations left behind when the photons separated from the primordial universe \cite{PLA2015}.  The full power spectrum of the CMB with covariance matrix is available in the latest \cosmomc.  In the next section, we use the full CMB power spectrum and covariance matrix to fit the cosmological parameters for the models we study.  Size limitations in the \cosmoejs package do not allow for use of the full CMB power spectrum with covariance matrix and likelihood code similar to Planck Legacy Archive, but does use the CMB distance priors as theoretical comparison parameters for the amplitude and locations of the acoustic peaks of the power spectrum.  We generate the CMB distance priors using \cosmomc and the full CMB power spectrum, following \cite{Wang2013}.  In the usual way, we use
the acoustic scale, $l_a$,  \cite{WangMukherjee,Wright},
\be
l_a=(1+z_*)\frac{\pi D_A(z_*)}{r_s(z_*)},
\label{eq:AcousticScale}
\ee
with the proper time angular diameter distance, $D_A(z_*)$ and the co-moving sound horizon, $r_s(z_*)$ as given earlier.  The redshift of the surface of last scattering of the CMB, $z_*$, is given by \cite{HuSugiyama}: 
\be
z_*=1048[1+0.00124(\Omega_b h^2)^{-0.738}][1+g_1 (\Omega_m h^2)^{g_2}].
\label{eq:zstar}
\ee
The constants $g_1$ and $g_2$ in the above expression are:
\be
g_1=\frac{0.0783(\Omega_b h^2)^{-0.238}}{1+39.5(\Omega_b h^2)^{0.763}},
\label{eq:zstarg1}
\ee
and
\be
g_2=\frac{0.560}{1+21.1(\Omega_b h^2)^{1.81}}.
\label{eq:ztarg2}
\ee
Finally, the shift parameter, $R$, \cite{Bond1997} is
\be
R(z_*)=\sqrt{\Omega_m}H_0(1+z_*)D_A(z_*).
\label{eq:ShiftParameter}
\ee
For comparing to data from the Wilkinson Microwave Anisotropy Probe (WMAP), these three parameters $x_i=\{l_a,R,z_*\}$ are used to generate a likelihood,  $\mathcal{L}_{CMB}=\triangle x_i$Cov$^{-1}(x_ix_j)\triangle x_j$ with $\triangle x_i=x_i-x^{obs}_i$ and Cov$^{-1}(x_ix_j)$ is the inverse covariance matrix for the parameters.  However, the authors in \cite{Wang2013}, modify this likelihood as  $x_i=\{\Omega_b h^2, l_a,R\}$.   We use the 2013 and 2015 releases of the Planck data sets \cite{PLA2013,PLA2015} to compute the distance priors for $x_i=\{\Omega_b h^2, l_a,R\}$, and the corresponding Cov$^{-1}$ from the full CMB TT, TE, EE and lowP power spectrum.  These are available for comparisons in \cosmoejs.  As an example, in the next sections we use the $\Lambda$CDM model as a background to build the CMB distance priors as seen in \cosmoejs.  Using \cosmomc, we compare the cosmological constraints of the $\Lambda$CDM model fit to the full CMB power spectrum and other data sets in \cosmomc with the same model fit to the distance priors covariance matrix and other data sets in \cosmomc.   In Figure \ref{fig:AllGRDEOmH0}, we can see the constraints on the cosmological parameters are tighter with the full CMB than with the CMB distance priors.  Also, there is overlapping parameter space for the mean values, which are statistically similar ($1 \sigma$) in most cases. These mean values will be used as initial parameter values to calculate the dynamics of models in \cosmoejs.  In table \ref{tb:constraintsII}, we provide the data set breakdown of our comparison of the use of the full CMB power spectrum and covariance matrix to that of the CMB distance priors.  This is expected as the CMB distance priors are derived from the full CMB power spectrum.  Finally, if we generate CMB distance priors using the full CMB power spectrum fit to the $\Lambda$CDM model, we will have a CMB background that we anticipate not to fit the DGP class of models \cite{IshakDEDGP2006, Dossett2010}.  This tension will affect the dynamics calculated with \cosmoejs using the fits of the DGP cosmological parameters.  In Sec. \ref{UsingCosmoMC}, we provide a robust example of a theoretical DGP model not matching the observational background of the CMB built from $\Lambda$CDM, and its dynamical consequences in Sec. \ref{UsingCosmoEJS}.  See for example, other CMB backgrounds that can be built using DE models \cite{HuangWang2015}.


\subsection{Alcock-Paczynski test}

The Alcock-Paczynski test determines the ratio of the radial (redshift) to the tangential (angular) size of objects assumed to be spherically symmetric.  This geometrical test is used to constrain the parameters of a particular model through the observable \cite{APOriginal1979},
\be
F(z) = \frac{\Delta z}{\Delta \theta} = (1+z)D_{A}(z) H(z) /c,
\label{eq:APtest}
\ee
which is constructed from the angular projection, $\Delta \theta = L_0/[(1+z)D_A(z)]$, and the radial projection, $\Delta z = L_0 H(z)/c$, with the assumption of an equal co-moving size, $L_0$.  \cosmoejs includes measurements of eq. (\ref{eq:APtest}) from \cite{APtestsSamuisha,PLA2015} and again compares AP data using the simple, $\chi^2_{AP}$, 

\begin{equation}
\chi^2_{AP}=\sum_{i=1}^{N}\frac{[F_{th}(z_i)-F_{obs}(z_i)]^2}{\sigma^2_i}.
\label{eq:APchisquare}
\end{equation}
Our latest version of \cosmomc utilizes the tabulated likelihoods described in \cite{APtestsSamuisha,PLA2015} to constrain the cosmological parameters of the models studied in the next section. 

\subsection{Growth factor parameter}

Complimentary constraints on cosmological models come from the growth rate of large scale structures of galaxy clusters, or redshift space distortions (RSD) \cite{SongPercivalRSD}.  When considering matter perturbations of linear order, the growth rate differential equation takes the following form, 
\be 
\ddot{\delta} + 2 H \dot{\delta} - 4\pi G_{eff} \rho_m \delta = 0,
\label{eq:Growthratedelta}
\ee
where $\delta = \delta \rho_m/ \rho_m$ is the matter density perturbation and $G_{eff}$ invokes the effect of modified gravity.  Transforming eq. (\ref{eq:Growthratedelta}) in terms of the logarithmic growth factor, $f = d \ln \delta / d \ln a$, we write
\be
f' + f^2 \Big(\frac{\dot{H}}{H^2}+2\Big)f = \frac{3}{2}\frac{G_{eff}}{G}\Omega_m,
\label{eq:Growthratef}
\ee
where $'$ is denotes $d/d\ln a$.  For comparisons to observations, we use the approximate form of the growth function $f$,
\be 
f = \Omega_m^{\gamma},
\label{growthfunction}
\ee
with $\gamma$ representing the growth index parameter of different cosmological models, i.e. $\gamma = 0.545$ for GRDE and $\gamma = 0.69$ for DGP.
 A particular model needs to satisfy both expansion history and growth of structures data sets.  In \cosmomc, a more robust comparison of $f(z)\sigma_8$ is used with likelihood methods \cite{PLA2015}.  We use the module likelihoods for the RSD contained in the latest version of \cosmomc.  The fiducial model chosen when measuring the redshift space distortions does affect the value of $f(z)\sigma_8$.  Since we are not expecting the DGP model to fit well, it is acceptable our data set assumes a $\Lambda$CDM fiducial cosmology.  \cosmoejs provides $f(z)\sigma_8$ comparison from
 \begin{equation}
\chi^2_{f}=\sum_{i=1}^{N}\frac{[X_{th}(z_i)-X_{obs}(z_i)]^2}{\sigma^2_i},
\label{eq:fchisquare}
\end{equation}
with $X(z_i)\equiv f(z_i),\,\textnormal{or}\,X(z_i)\equiv f(z_i)\sigma_8$, where $X(z_i)\equiv f(z_i)$ is only for older $f(z)$ data sets, where a $f(z) \sigma_8$ is not provided.  In our next sections we do not use any $f(z)$-only data sets in the fitting or dynamics.

 
 \section{Results from using \cosmomc \label{UsingCosmoMC}}
 
\begin{table}

\begin{tabular}{|c|c|c|c|c|c|c|} \hline
$\Lambda$CDM & CMB & Full &DGP &CMB & DGP $+k$ &CMB	 \\ 
Model& distance& CMB& Model & distance & Model & distance	 \\ \hline
$\chi^2_{Total}$ & $731.80$&$12381.00$&$\chi^2_{Total}$ & $907.27$&$\chi^2_{Total}$ & $768.25$ \\ \hline
$\chi^2_{CMB}$ &$1.24$  &$12964.00$ &$\chi^2_{CMB}$ &$43.07$  &$\chi^2_{CMB}$ &$1.18$ \\ \hline
$\chi^2_{JLA}$ & $695.15$&$695.16$ &$\chi^2_{JLA}$ & $728.93$ &$\chi^2_{JLA}$ & $711.09$\\ \hline
$\chi^2_{H_0}$ & $0.08$&$0.08$ &$\chi^2_{H_0}$ & $16.80$ &$\chi^2_{H_0}$ & $4.03$\\ \hline
$\chi^2_{H(z)}$ & $25.29$&$25.27$ &$\chi^2_{H(z)}$ & $66.87$ &$\chi^2_{H(z)}$ & $37.59$\\ \hline
$\chi^2_{BAO+RSD}$ & $9.28$&$9.45$&$\chi^2_{BAO+RSD}$ & $50.58$ &$\chi^2_{BAO+RSD}$ & $14.35$\\ \hline
$H_0$ & $67.72^{+0.80}_{-0.78}$&$67.70^{+0.86}_{-0.84}$ &$H_0$ & $56.57^{+0.59}_{-0.60}$&$H_0$ & $63.575^{+1.340}_{-1.341}$\\ \hline
$\Omega_m$ &$0.305^{+0.01}_{-0.01}$&$0.303^{+0.01}_{-0.01}$&$\Omega_m$ &$0.397^{+0.01}_{-0.01}$ &$\Omega_m$ &$0.316^{+0.015}_{-0.015}$ \\ \hline
Age (Gyr) &13.86 &13.77 &Age (Gyr) &14.58 &Age (Gyr) &13.72 \\ \hline

\end{tabular}

\caption{Best-fit $\chi^2$ likelihoods for the $\Lambda$CDM model from the GRDE class using eq. (\ref{eq:FriedmannEquationH}), the flat DGP model of eq. (\ref{eq:FriedmannDGP}), and the curved DGP$+k$ model with comparisons to the CMB distance priors derived from the full 2015 Planck TT, TE, EE and lowP data release \cite{PLA2015} to the full CMB with covariance matrix.  Both fits use the same additional data sets from Sec. \ref{DatainCosmoEJS}: $H_0$ locally from Cephied variables \cite{H070p6}, $H(z)$ measurements as given in Table \ref{Hzdata}, supernovae from the JLA compilation \cite{JLA},  BAO from 6dFGS \cite{Beutler6dFGS} , MGS \cite{RossMGS}, and  DR12 BOSS CMASS \cite{DR12BOSSCMASS}, DR12 BOSS LOWZ \cite{DR12BOSSLOWZ} with RSD measurements (which include AP and $f(z)\sigma_8$ measurements).  }
\label{tb:constraintsII}
\end{table}

\begin{table}
\resizebox{5.8in}{!} {
\begin{tabular}{|c|c|c|c|c|c|c|} \hline
Model& $H_0$& $\Omega_m$	&$\Omega_k$ &$w$ &$w_a$&$\chi^2$ \\ \hline
$\Lambda$CDM&$67.717^{+0.801}_{-0.781}$  &$0.305^{+0.010}_{-0.009}$ &- &- &-&731.8 \\ \hline
$\Lambda$CDM$+k$& $68.426^{+1.484}_{-1.446}$&$0.299^{+0.014}_{-0.013}$ &$0.001^{+0.002}_{-0.002}$ &- &-&730.4 \\ \hline

$w$CDM&$67.555^{+1.881}_{-1.811}$ &$0.306^{+0.015}_{-0.015}$ &- &$-0.994^{+0.057}_{-0.060}$ & -&732.0\\ \hline

$w$CDM$+k$&$67.648^{+1.881}_{-1.845}$ &$0.304^{+0.015}_{-0.014}$ &$0.002^{+0.003}_{-0.002}$ &$-0.951^{+0.076}_{-0.074}$&-&729.2 \\ \hline

CPL &$67.265^{+1.878}_{-1.833}$ &$0.310^{+0.016}_{-0.017}$ &- &$-0.873^{+0.160}_{-0.166}$&$-0.436^{+0.488}_{-0.564}$&729.8 \\ \hline

CPL$+k$&$67.582^{+2.127}_{-2.053}$ &$0.305^{+0.020}_{-0.021}$ &$0.002^{+0.004}_{-0.003}$ &$-0.926^{+0.188}_{-0.202}$&$-0.123^{+0.727}_{-0.803}$&729.6 \\ \hline

DGP&$56.566^{+0.586}_{-0.595}$ &$0.397^{+0.011}_{-0.011}$ &- &- &-&907.2 \\ \hline

DGP$+k$&$63.575^{+1.340}_{-1.341}$ &$0.316^{+0.015}_{-0.015}$ &$0.014^{+0.002}_{-0.002}$ &- &-&768.2 \\ \hline

\end{tabular}
}
\caption{$95\%$ confidence constraints on parameters and best-fit $\chi^2$ likelihoods for models from the GRDE class using eq. (\ref{eq:FriedmannEquationH}) and the DGP class given eq. (\ref{eq:FriedmannDGP}) from comparisons to the CMB distance priors derived from the full 2015 Planck TT, TE, EE and lowP data release \cite{PLA2015}, $H_0$ locally from Cephied variables \cite{H070p6}, $H(z)$ measurements as given in Table \ref{Hzdata}, supernovae from the JLA compilation \cite{JLA},  BAO from 6dFGS \cite{Beutler6dFGS} , MGS \cite{RossMGS}, and  DR12 BOSS CMASS \cite{DR12BOSSCMASS}, DR12 BOSS LOWZ \cite{DR12BOSSLOWZ} with RSD measurements (which include AP and $f(z)\sigma_8$ measurements).  For all the models listed in this Table, `+k' models allow curvature to be constrained by the data, and those absent `+k' fix $\Omega_k=0.0$.  Specifically, the  `CPL+k' and `CPL' models use the full eq. (\ref{eq:FriedmannEquationH}); both the `$w$CDM+k' and `$w$CDM' models fix $w_a=0.0$ in eq. (\ref{eq:FriedmannEquationH}); and finally, the `$\Lambda$CDM+k' and `$\Lambda$CDM' models allow only a cosmological constant, $\Lambda$, dark energy, so $w=-1.0$ and $w_a=0.0$. }
\label{tb:constraints}
\end{table}

\begin{figure}[h!]
\begin{center}
\begin{tabular}{|c|c|}\hline$+k$

\title{$\Lambda$CDM, $\Lambda$CDM$+k$, $w$CDM, $w$CDM$+k$, CPL and CPL$+k$ comparing $(\Omega_m,\, H_0)$}

\includegraphics[scale=0.45]{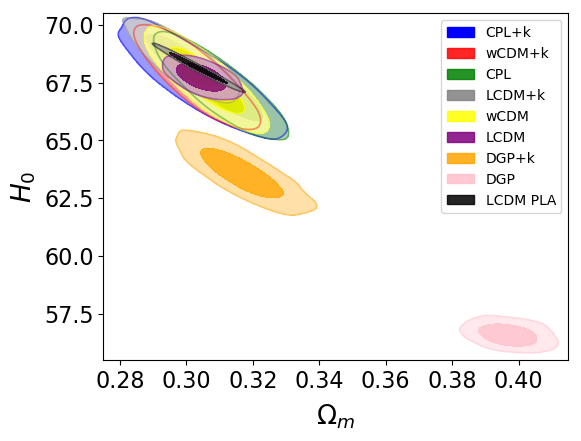} &  \includegraphics[scale=0.45]{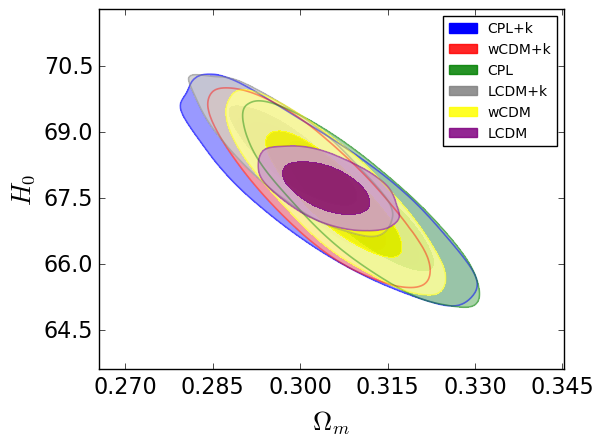}\\
\hline
\end{tabular}

\caption{(color online) Left: 2D contour plot showing $68\%$ and $95\%$ confidence limits on $(\Omega_m,\, H_0)$  for the two DGP models, `DGP' and `DGP+k' constructed from eq.  (\ref{eq:FriedmannDGP}) for flat (pink) $(\Omega_k = 0)$ and curved (orange) cases, respectively and for the six models constructed from eq. (\ref{eq:FriedmannEquationH}) when constrained by all the observations given in Sec. \ref{DatainCosmoEJS}.  Models based on GR and include a dark energy component, which in some cases, may be a cosmological constant, $\Lambda$, with $w=-1.0$ and $w_a=0.0$, for the `$\Lambda$CDM' (purple) and `$\Lambda$CDM+k' (grey) models.  The models missing `+k' fix $\Omega_k=0.0$, the `$w$CDM' (yellow) and `$w$CDM+k' (red) model allow different dark energy equations of state, but do not vary, $w_a=0.0$, and the `CPL' (green) and `CPL+k' (blue) models allow fitting of $w_a$. Finally, the $\Lambda$CDM (black) model tested with the full 2015 Planck data is provided for comparison.  Right: Same as left, but without the DGP models. }
\label{fig:AllGRDEOmH0}

\end{center}
\end{figure}

\begin{figure}[h!]
\begin{center}
\begin{tabular}{|c|c|}\hline
\title{$\Lambda$CDM$+k$, $w$CDM$+k$, CPL$+k$ and DGP$+k$ comparing $(\Omega_k,\, H_0)$;$w$CDM, $w$CDM$+k$, CPL and CPL$+k$ comparing $(w,\, H_0)$}

\includegraphics[scale=0.45]{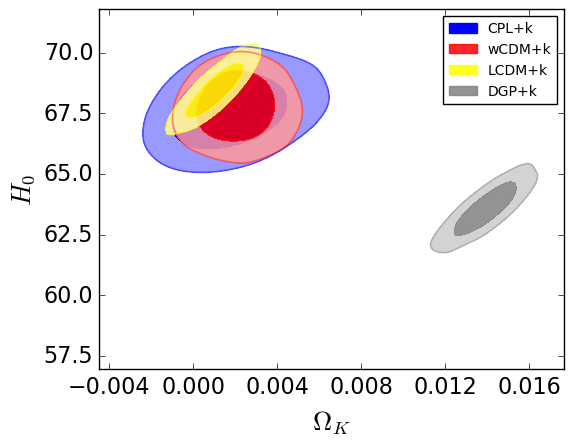} & \includegraphics[scale=0.45]{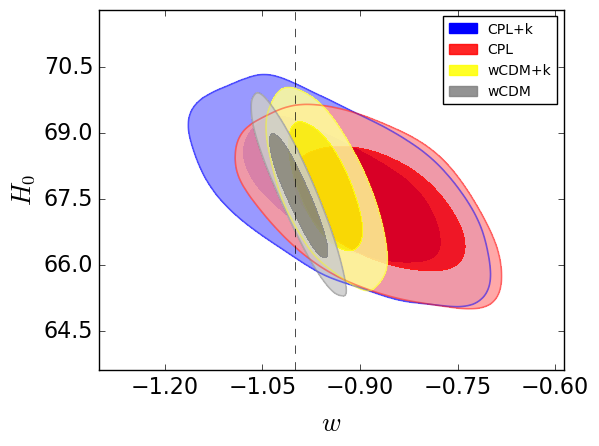}\\
\hline
\end{tabular}
\caption{(color online) 2D contour plots showing $68\%$ and $95\%$ confidence limits on $(\Omega_k,\, H_0)$, and $(w,\, H_0)$ with all the observations in Sec. \ref{DatainCosmoEJS}, left and right, respectively.  Left: See text for the description of three models constructed from eq. (\ref{eq:FriedmannEquationH}) with $\Omega_k$ and the fourth model, `DGP+k' (grey) following eq. (\ref{eq:FriedmannDGP}), and requires an open universe, $\Omega_k>0$, or negatively curved, $k=-1$ spacetime, see eq. (\ref{eq:Curvaturedensity}). Right: The four models constructed from eq. (\ref{eq:FriedmannEquationH}) that allow a variant of the dark energy equation of state parameter, $w$.  While `+k' signifies models which allow fitting of $\Omega_k$, the `CPL' (red) and `CPL+k' model vary $w_a$, the derivative of the equation of state, as well, but the `$w$CDM' (grey) and `$w$CDM+k' (yellow) models fix $w_a=0.0$.}
\label{fig:GROkH0wH0}

\end{center}
\end{figure}
\begin{figure}[h!]
\begin{center}
\begin{tabular}{|c|c|}\hline
\title{CPL and CPL$+k$ comparing $(w,\, w_a)$; DGP and DGP$+k$ comparing $(\Omega_m,\, H_0)$}

\includegraphics[scale=0.45]{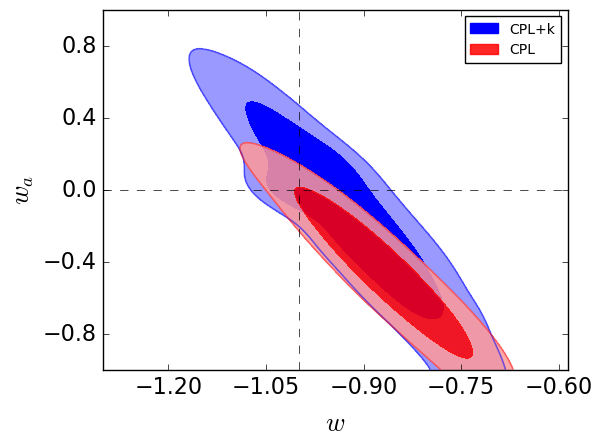} & \includegraphics[scale=0.45]{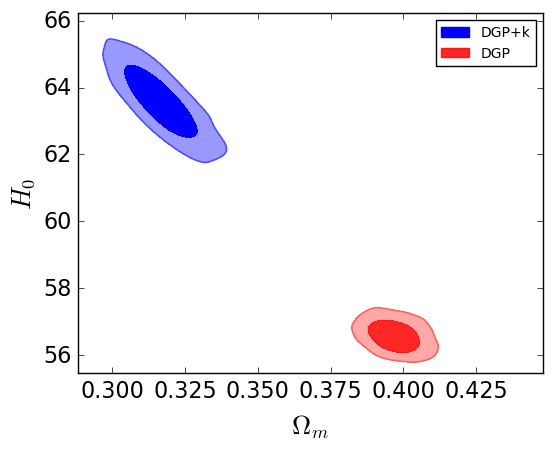}\\
\hline
\end{tabular}
\caption{(color online) Left: 2D contour plot showing $68\%$ and $95\%$ confidence limits on $(w,\, w_a)$ for the two models, `CPL' and `CPL+k', constructed from eq. (\ref{eq:FriedmannEquationH}) for flat, $\Omega_k=0.0$, (red) and curved (blue) cases, respectively, when comparing to the observations of Sec. \ref{DatainCosmoEJS}.  Right: 2D contour plot showing $68\%$ and $95\%$ confidence limits on $(\Omega_m,\, H_0)$ for the two DGP models, `DGP' and `DGP+k' constructed from eq.  (\ref{eq:FriedmannDGP}) for flat (red) $(\Omega_k = 0)$ and curved (blue) cases, respectively, when comparing to the observations of Sec. \ref{DatainCosmoEJS}.}
\label{fig:GRw0waDGPOmH0}

\end{center}
\end{figure}

We constrain the free parameters of six models built from the GRDE class eq. (\ref{eq:FriedmannEquationH}) and two models built from the DGP class eq. (\ref{eq:FriedmannDGP}) using \cosmomc, a Markov chain Monte Carlo program, with combinations of observations given in Sec. \ref{DatainCosmoEJS} using covariance matrix $\chi^2$ methods. Specifically, we compare the models to the CMB distance priors derived from the full 2015 Planck TT, TE, EE and lowP data release \cite{PLA2015}, $H_0$ locally from Cephied variables \cite{H070p6}, $H(z)$ measurements as given in Table \ref{Hzdata}, supernovae from the JLA compilation \cite{JLA},  BAO from 6dFGS \cite{Beutler6dFGS} , MGS \cite{RossMGS}, and  DR12 BOSS CMASS \cite{DR12BOSSCMASS}, DR12 BOSS LOWZ \cite{DR12BOSSLOWZ} with RSD measurements (which include AP and $f(z)\sigma_8$ measurements).  Briefly, we give \cosmomc a range of priors for initial values of the model parameters and it returns the fits of these parameters for a particular model tested.  

The fits we obtain are shown in Table \ref{tb:constraints}.  In each model, `+k' identifies a model that allows fitting of the curvature density parameter, $\Omega_k$, and those without `+k' use $\Omega_k=0.0$.  The `$\Lambda$CDM' and `$\Lambda$CDM+k' models hold the equation of state parameter, $w=-1.0$ and its derivative, $w_a=0.0$, yielding a cosmological constant dark energy equation of state. The `$w$CDM' and `$w$CDM+k' models additionally fix $w_a=0.0$, and allow the dark energy equation of state, $w$ to be fit by the data.  The more general `CPL' and `CPL+k' models let the data constrain both $w$ and its derivative, $w_a$.  

A lower $\chi^2$ value in Table \ref{tb:constraints} corresponds to a better fit to the data, however, in the cases of the GRDE models, the lower $\chi^2$ is primarily due to the extra degrees of freedom allowed to the model by additional parameters allowing a lower $\chi^2$.  The constraints we obtain on the dark energy class of models are consistent with recent cosmological fits found elsewhere in the literature \cite{PLA2015}.  As expected, due to the $\Lambda$CDM CMB distance priors background, the flat ($\Omega_k=0$) DGP model and curved DGP model have the worst fits to all the data sets, where the improved fit of the curved DGP comes from the extra parameter space, i.e. $\Omega_k$.  From these results we choose the $\Lambda$CDM, curved DGP$+k$ and flat DGP models as the statistically most preferred and least preferred models from each class to dynamically simulate in the next section with \cosmoejs.  We also provide the breakdown of the $\chi^2$ for each data set in Table \ref{tb:constraintsII} for the models we study in Sec. \ref{UsingCosmoEJS}.

 We provide some of the 2D contour plots for parameters of interest in Figures \ref{fig:AllGRDEOmH0}, \ref{fig:GROkH0wH0}, \ref{fig:GRw0waDGPOmH0}, and we use these range of values to illustrate the dynamical analysis of the models in Sec. \ref{UsingCosmoEJS}.  We see evidence in Figure \ref{fig:AllGRDEOmH0}, that a more general GRDE model, such as the `CPL+k' model relaxes the constraint on $H_0$ because of the added freedom with the parameters, $\{w,\,w_a,\,\Omega_k$\}, as seen in \cite{H0Riess2016}.  In Figures \ref{fig:GROkH0wH0}, \ref{fig:GRw0waDGPOmH0} (left), we have the $68\%$ and $95\%$ confidence contours show a combination of ranges for $\{\Omega_k,\,H_0\}$, $\{w,\,H_0\}$, $\{w,\,w_a\}$, respectively.  As expected, the DGP models have a more serious tension with the data sets and even each other, as seen in Figure \ref{fig:GRw0waDGPOmH0} (right), because of the constructed CMB distance priors from the $\Lambda$CDM background.

\section{Exploring the dynamics of cosmological models using \cosmoejs \label{UsingCosmoEJS}}

Considering the fits from Table \ref{tb:constraints}, it is clear to see the observationally favored models from the $\chi^2$ values and their respective free parameter means and standard deviations.  However, how these fits were achieved, requires some explanation.  Not how does MCMC work, but which data points had tension with the cosmological model causing increased $\chi^2$ values?  Does the model fit equally well to low and high redshift data points, and how does the precision of each affect the fitting of the models.  The dynamical plots of \cosmoejs help to answer these questions.

 After achieving the fits for the free parameters of a particular model in Sec. \ref{UsingCosmoMC}, we use \cosmoejs to reproduce the dynamics of the models for those parameter values.  As we describe, this new version of \cosmoejs simulates the theoretical dynamics of two different classes of cosmological models, GRDE and DGP, simultaneously while comparing them to the latest observational data sets.   \cosmoejs is a package of \texttt{Java} simulations and modeling programs for cosmology built from Easy Java Simulations (EJS).  EJS is a Java-based software package that combines the high performance coding language of Java with easy-to-use graphical user interfaces (GUIs) and real-time plotting for building interactive modeling simulations \cite{EJS}.  The \cosmoejs package contains different modules depending on which observations and models are under study.  For a complete description of the usage of \cosmoejs we refer the reader to the supplemental documents of the package and \cite{CosmoEJSPaper}.  Briefly, a Java GUI opens with a tabbed plot frame.  Sliders allow the user to select parameter values for the model and drop-down menus contain data sets to choose to compare with the model.  Based on the parameter values chosen, a specific model's theoretical simulated data is simultaneously compared to selected data sets of actual cosmological observations.  The programs contain fitting methods as described in Sec. \ref{DatainCosmoEJS} to numerically prefer one set of parameter values over another set, however, as we have already found the preferred values in Sec. \ref{UsingCosmoMC}, we focus on the visual inspection of the model's dynamics from the plots generated by \cosmoejs.  

In \cite{CosmoEJSPaper}, the dynamics of the $\Lambda$CDM model were studied with \cosmoejs, where particular attention was paid to visual differences in the plots showing the fits of the models to recent data sets.  To demonstrate the benefit of having \cosmoejs as a dynamical tool for a broader collection of models, we highlight the dynamics of the flat DGP model and the more competitive curved DGP$+k$ model to compare them to the $\Lambda$CDM model.  One of these models ($\Lambda$CDM) was the statistically favored model (low $\chi^2$) from the two classes in Sec. \ref{UsingCosmoMC} and the flat DGP was clearly the least statistically favored model.  \cosmoejs can accurately demonstrate the dynamics of all the models shown in Table \ref{tb:constraints}, but the flat DGP model provides a clear example from which to emphasize the visual perspective of its dynamics and the curved DGP$+k$ model shows that more competitive models have less visual differences and require statistical methods to find preferred fits to precise data.  
\begin{figure}[h!]
\begin{center}
\begin{tabular}{|c|}\hline
\title{DE and DGP CMB and Summary}
\bf DE CMB Output\\
\includegraphics[scale=0.42]{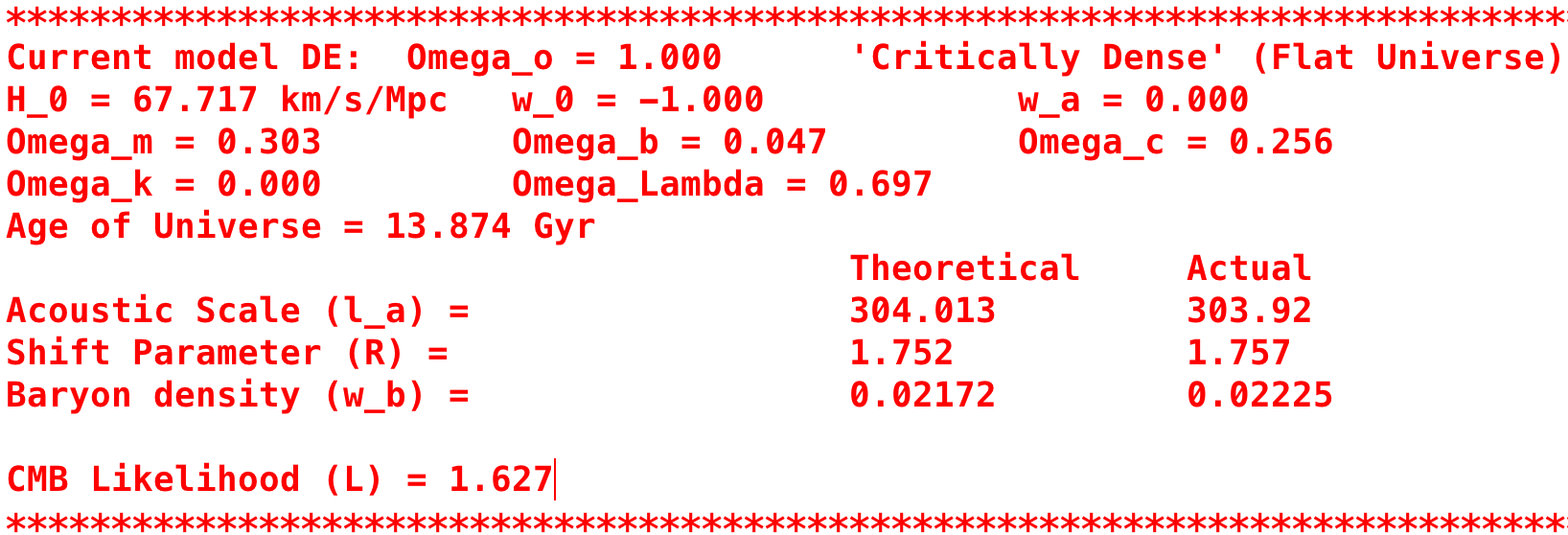}\\
\hline
\bf DGP CMB Output\\
\includegraphics[scale=0.42]{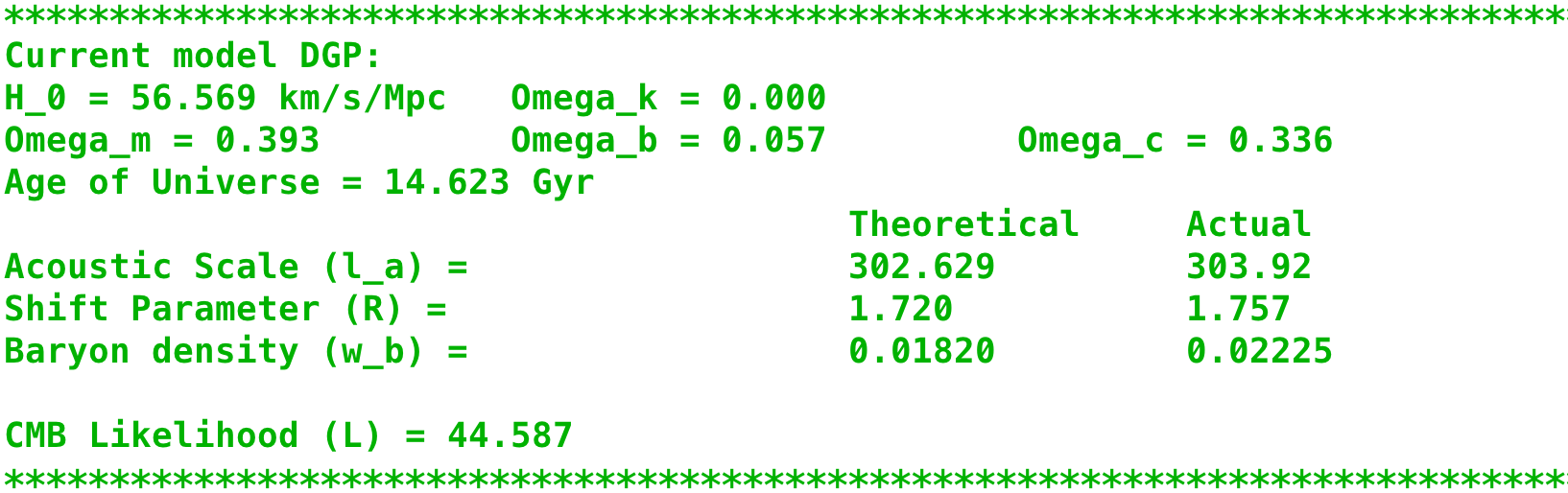}\\
\hline
\bf DGP$+k$ CMB Output\\
\includegraphics[scale=0.42]{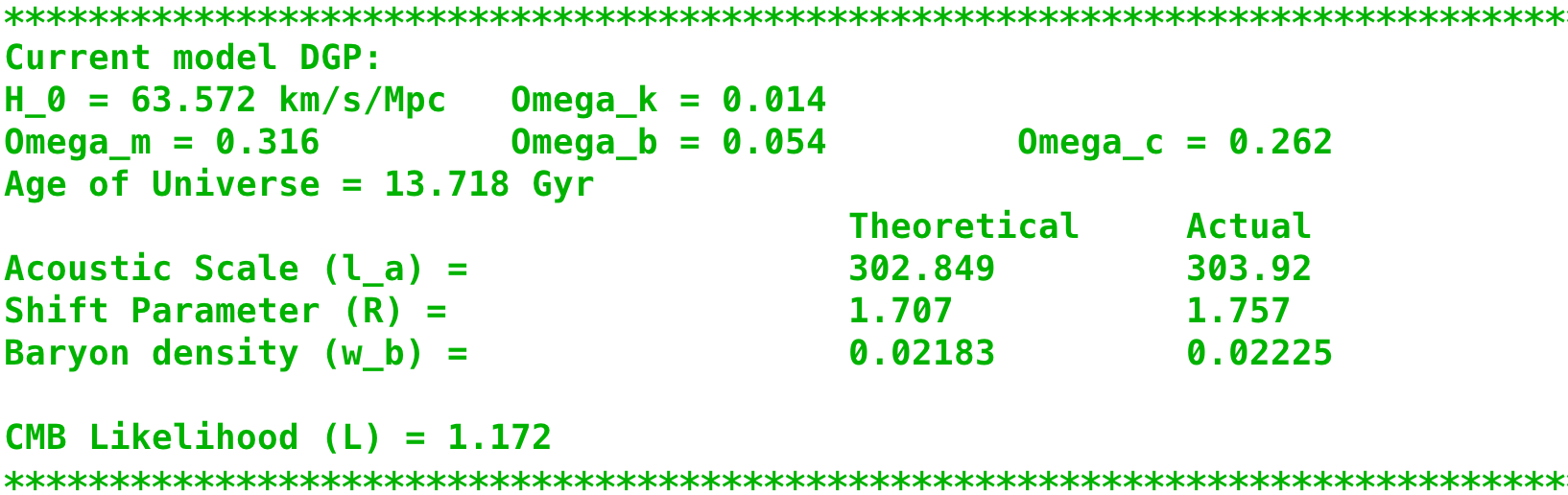}\\
\hline
\end{tabular}
\caption{(color online)  The CMB output panel summarizes all the initial parameter values  for the $\Lambda$CDM model (top), the flat DGP model (middle), and the curved DGP$+k$ model (bottom), (color online matches the model) used in the calculation for comparison to the observational data on the other panels (see Figure \ref{fig:CosmoEJS}), as well as, comparing the model to the CMB last scattering surface data.  Clearly, the $\Lambda$CDM model and the curved DGP$+k$ model have a lower $\chi^2$ and a better fit to the CMB data. }
\label{fig:DEDGPCMB}

\end{center}
\end{figure}

\begin{figure}[h!]
\begin{center}
\begin{tabular}{|c|c|}\hline
\title{CosmoEJS}
\includegraphics[scale=0.24]{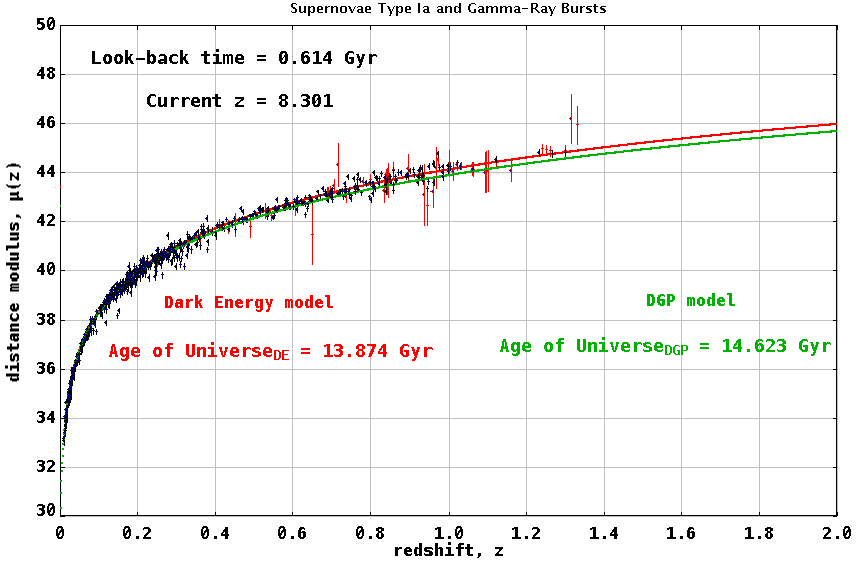} & \includegraphics[scale=0.24]{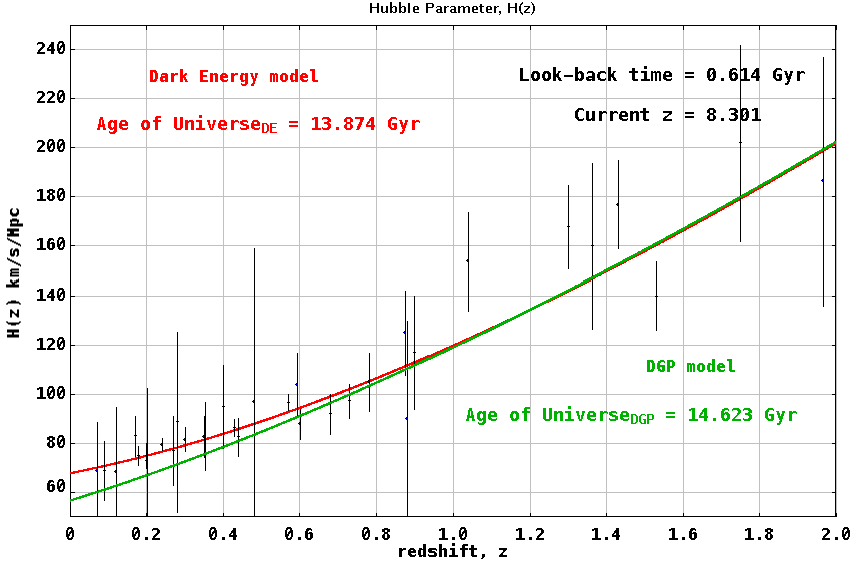}\\
 \hline
\includegraphics[scale=0.24]{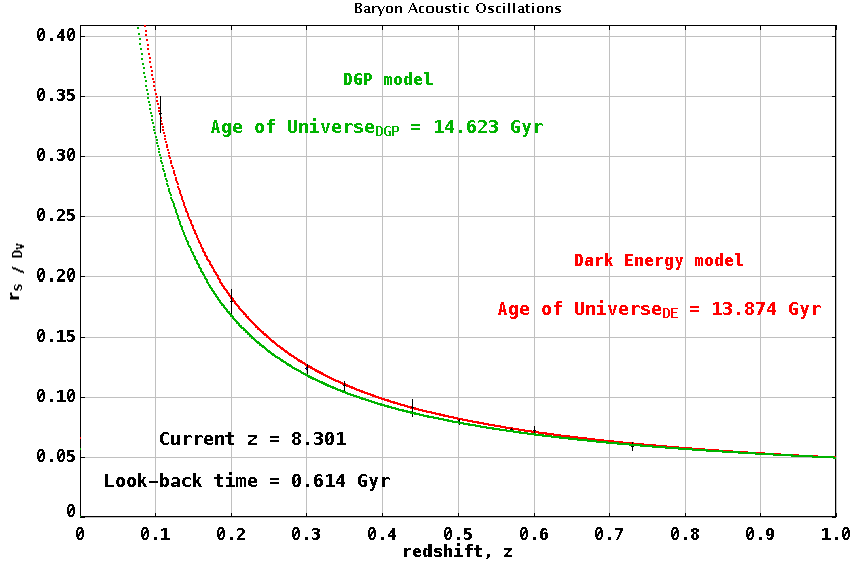} & \includegraphics[scale=0.24]{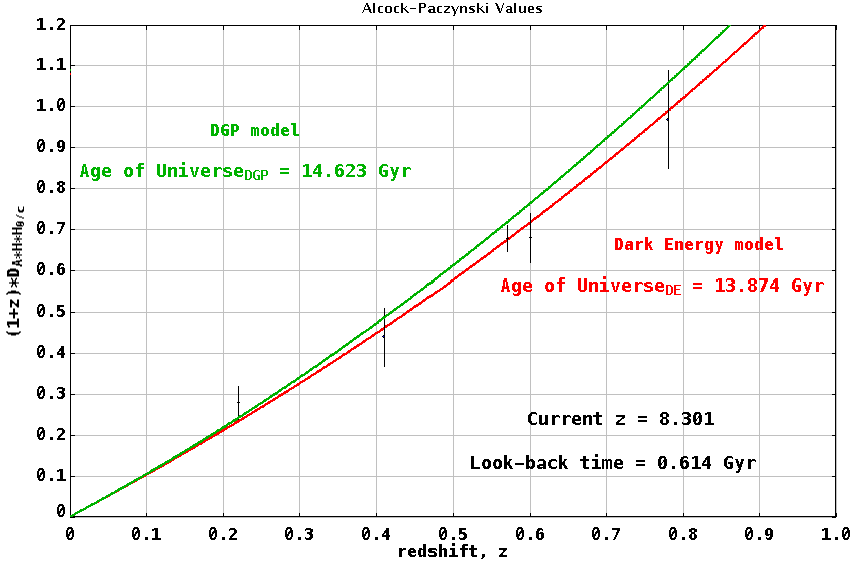}\\ 
\hline
\includegraphics[scale=0.24]{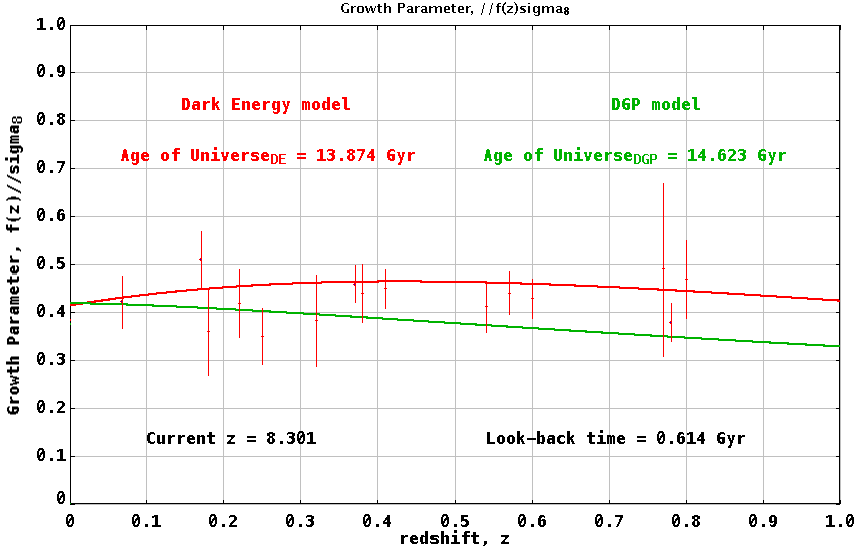} & \includegraphics[scale=0.24]{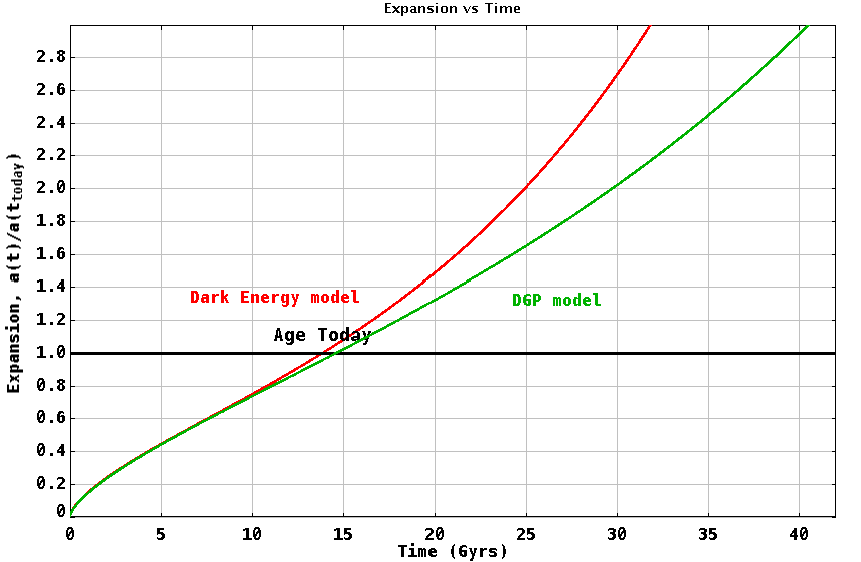}\\

\hline
\end{tabular}
\caption{A graphical and numerical (Age) comparison of the best-fit $\Lambda$CDM and flat DGP models in Table \ref{tb:constraints} to actual observational data of the supernovae type Ia and gamma ray bursts (top left), the Hubble Parameter $H(z)$ (top right), the baryon acoustic oscillations (middle left), the Alcock-Paczynski test (middle right), growth factor parameter $f(z) \sigma_8$ (bottom left) and the expansion history (bottom right) from several surveys using \cosmoejs.  The program simultaneously calculates simulated data for each model for the initial values of the model's parameters and plots them for fitting of the data and visual inspection. See text for more comparison details with these data sets. }
\label{fig:CosmoEJS}

\end{center}
\end{figure}

\begin{figure}[h!]
\begin{center}
\begin{tabular}{|c|c|}\hline
\title{CosmoEJS}
\includegraphics[scale=0.24]{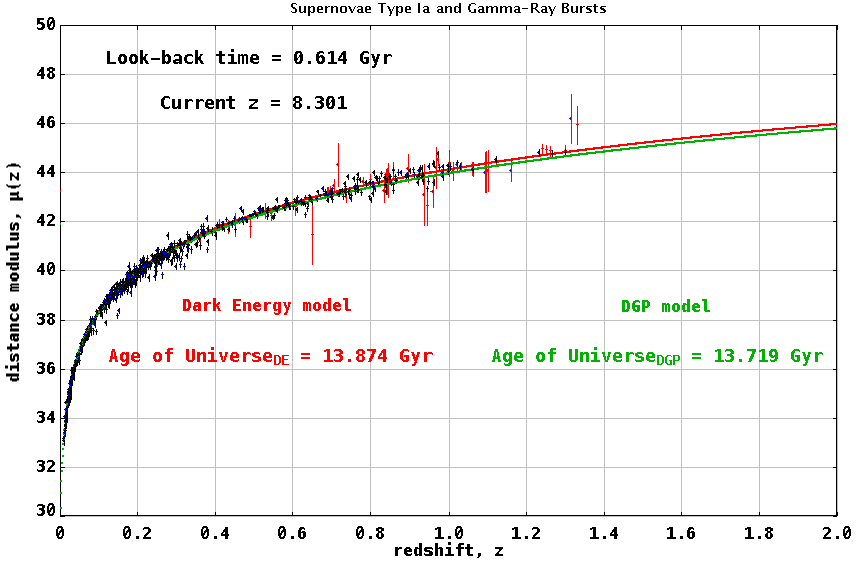} & \includegraphics[scale=0.24]{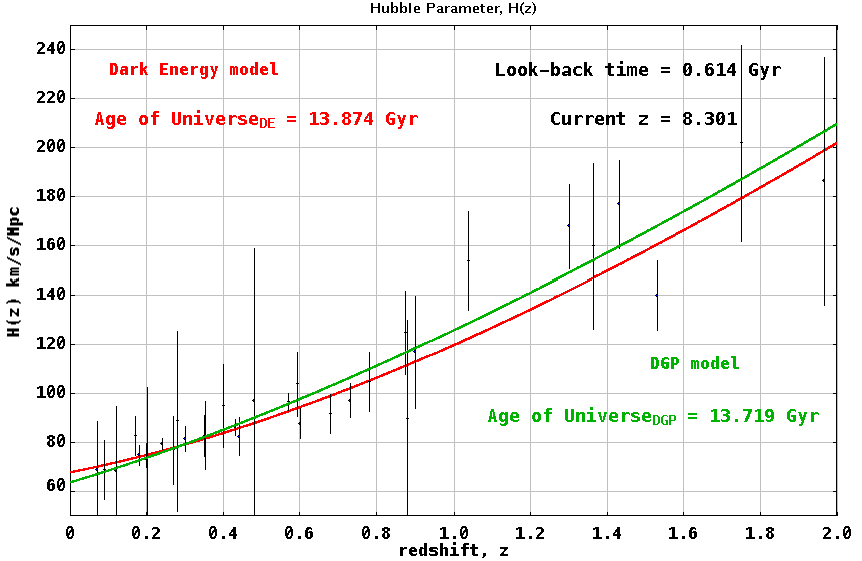}\\
 \hline
\includegraphics[scale=0.24]{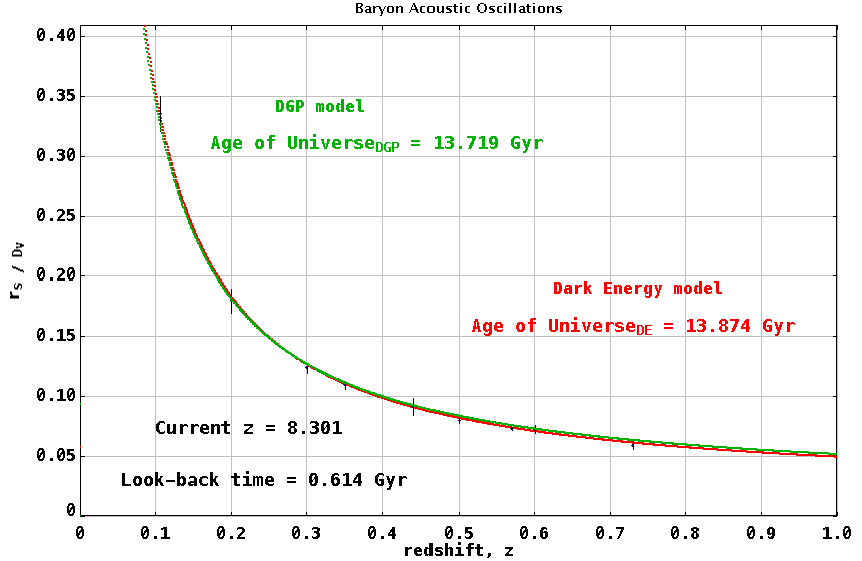} & \includegraphics[scale=0.24]{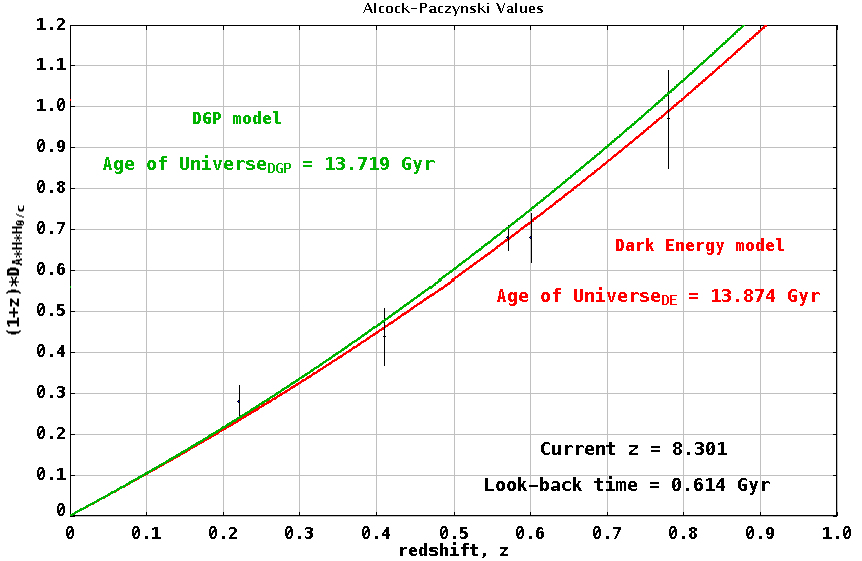}\\ 
\hline
\includegraphics[scale=0.24]{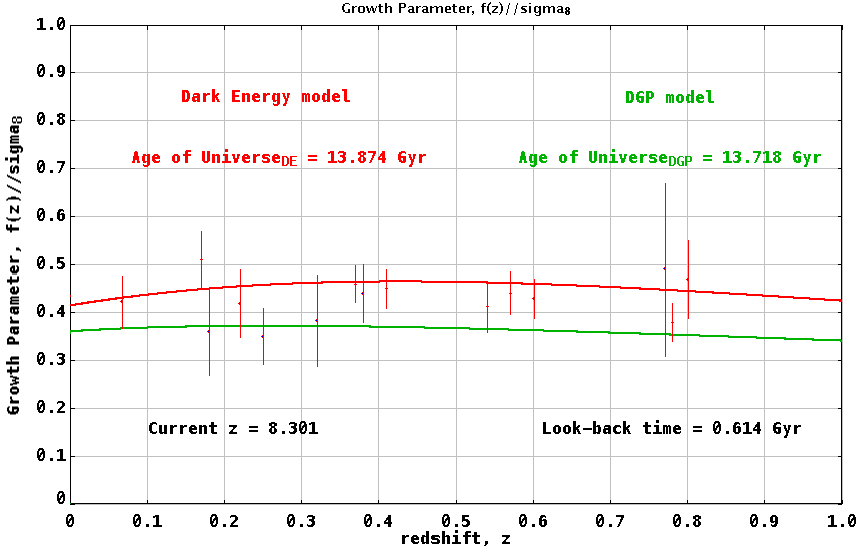} & \includegraphics[scale=0.24]{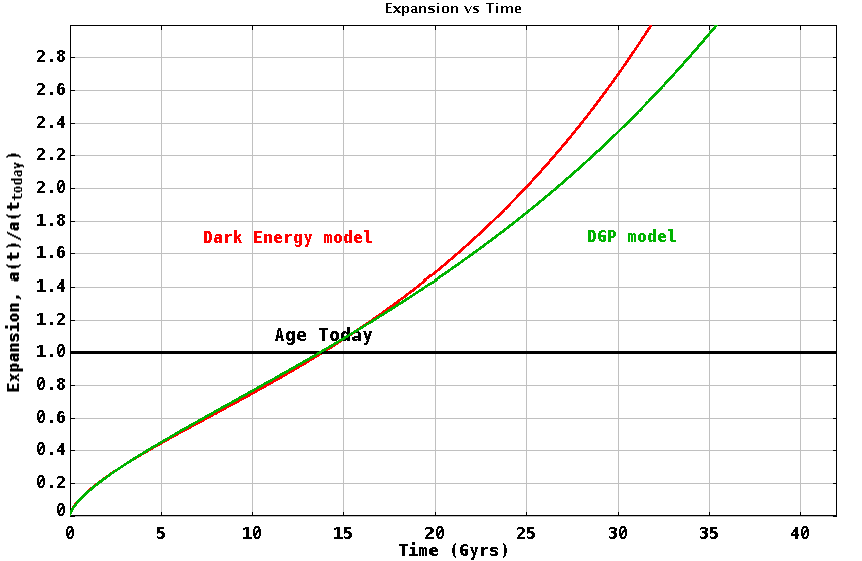}\\

\hline
\end{tabular}
\caption{A graphical and numerical (Age) comparison of the best-fit $\Lambda$CDM and DGP$+k$ models in Table \ref{tb:constraints} to actual observational data of the supernovae type Ia and gamma ray bursts (top left), the Hubble Parameter $H(z)$ (top right), the baryon acoustic oscillations (middle left), the Alcock-Paczynski test (middle right), growth factor parameter $f(z) \sigma_8$ (bottom left) and the expansion history (bottom right) from several surveys using \cosmoejs.  The program simultaneously calculates simulated data for each model for the initial values of the model's parameters and plots them for fitting of the data and visual inspection. See text for more comparison details with these data sets. }
\label{fig:CosmoEJSk}

\end{center}
\end{figure}

\begin{figure}[h!]
\begin{center}
\begin{tabular}{|c|c|}\hline
\title{CosmoEJS}
\includegraphics[scale=0.24]{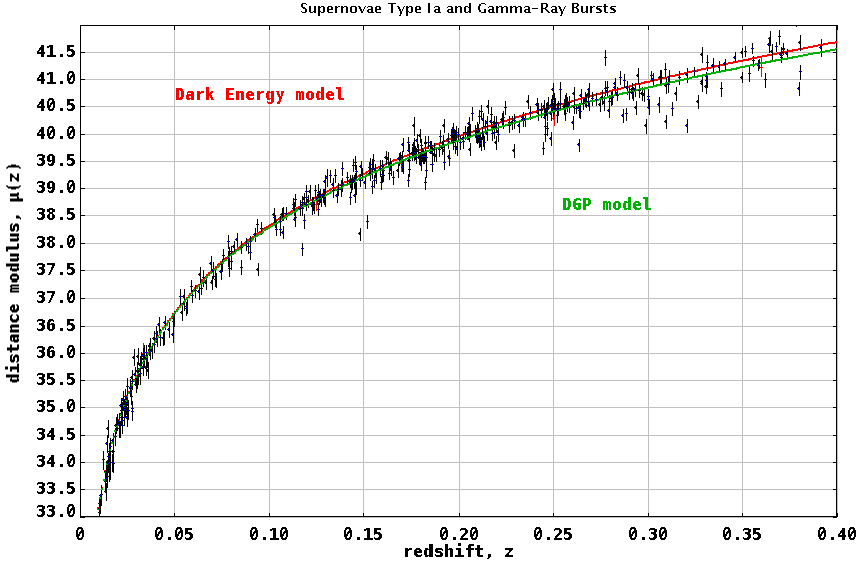} & \includegraphics[scale=0.24]{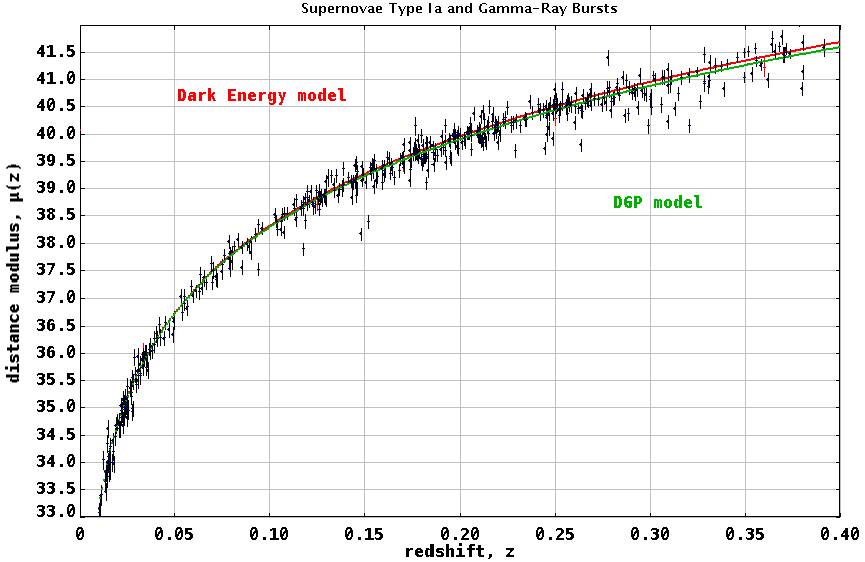}\\

\hline
\end{tabular}
\caption{A low redshift comparison of SNeIa to the best-fit $\Lambda$CDM model, flat DGP model (left) and curve DGP$+k$ model (right) in Table \ref{tb:constraints} using \cosmoejs.  This closer inspection of Figures \ref{fig:CosmoEJS} and \ref{fig:CosmoEJSk} does show a difference between the models, but it is unclear from these plots which is the preferred model.  The supernovae data are more numerous and more precise at low redshift, so the graphical comparison with \cosmoejs shows that this is an important area to explore and fit statistically with \cosmomc using a covariance matrix method.   }
\label{fig:CosmoEJSzoom}

\end{center}
\end{figure}

We use the mean values from Table \ref{tb:constraints} to set initial values for each of the parameters in the $\Lambda$CDM model of equation (\ref{eq:FriedmannEquationH}), the curved DGP$+k$ and the flat DGP models of equation (\ref{eq:FriedmannDGP}).  \cosmoejs simultaneously calculates the theoretical values for the observables described in Sec. \ref{DatainCosmoEJS} for both models and then compares them graphically and numerically to the data sets selected.  A summary of the parameter values chosen for the models to compare to the observations in the \cosmoejs graphing panels are provided in the CMB panels shown in Figure \ref{fig:DEDGPCMB}, as well as, a comparison of the theoretical models to the CMB distance priors.   
\footnote{The $\chi^2$ calculation for other observables in \cosmoejs is merely a comparative minimum chisquare method and does not use a covariance matrix, except in the case of the CMB distance priors which is built from results of \cite{PLA2015}.  These $\chi^2$ values are not used to fit the model in \cosmoejs.  As seen in \cite{CosmoEJSPaper}, because this minimum chisquare method weights all data equally, some models are preferred when they should not be.} From the theoretical values in the CMB panel, i.e. $\{\omega_b, l_a,R\}$ ($\omega_b = \Omega_b h^2$), we can directly see why the flat DGP model has a poorer fit to the data than the $\Lambda$CDM model.  

We see similar comparisons with visual inspection of the dynamical plots on the SNeIa and GRB, $H(z)$, BAO, AP and Growth parameter panels in Figures \ref{fig:CosmoEJS} and \ref{fig:CosmoEJSk}. The simultaneous plotting of both models allows for real-time comparisons of different classes of models to all the data sets for complimentary fits.  
We do not show the $\chi^2$ in the plotting panels from \cosmoejs as it is the general minimum chisquare method built to accommodate various data sets and are not accurate to the covariance matrix method used \cosmomc in Sec. \ref{UsingCosmoMC}.  For more similar models, only the covariance matrix method should be used.  When the visual differences are less prominent, it is more important to use the covariance matrix fitting methods.  All further references to $\chi^2$ in this section refer to covariance matrix methods used in \cosmomc and the analysis of Sec. \ref{UsingCosmoMC}. 

Considering Figure \ref{fig:CosmoEJS} (top left), we see evidence of the cosmic acceleration.  Nearby SNeIa redshift values are larger than they should be for a constant expansion and higher redshift SNeIa are not as far away, either.  While each model uses a different physical mechanism (see Sec. \ref{models}), both fit the cosmic acceleration evidenced by SNeIa and GRB, similarly, see Figure \ref{fig:CosmoEJS}.  Now with graphical aid of \cosmoejs, the difference between the models appears to be more pronounced at higher redshift.  Visually, at low redshift the models are very similar.  Upon a closer inspection of the SNeIa graphs in Figure \ref{fig:CosmoEJSzoom}, we do see a difference between the models, but it is not visually clear which model is preferred.    This further emphasizes the need for a covariance matrix approach to properly prefer one model to another.  A more rigorous method of checking these low redshift SNeIa would involve refitting this subset of supernovae using a covariance matrix for the same subset.  Similarly,  in the $H(z)$ Hubble Parameter panel, Figures \ref{fig:CosmoEJS} (top right) and \ref{fig:CosmoEJSk} (top right), the different input values of the Hubble Constant, $H(z=0)$ for the $\Lambda$CDM model, the curved DGP$+k$ model and the flat DGP model are easily seen.  Again, graphically, we also observe differences in the shape of their curves leading to the higher $\chi^2$ for the flat DGP model as it deviates from the more precise $H(z)$ data points at lower redshift, while all models have similar trajectories near less precise higher redshift data points.  

A large deviation in $\chi^2$ values between the flat DGP model and the $\Lambda$CDM model comes from the BAO data in Figure \ref{fig:CosmoEJS} (middle left).  Visually, both models have the same shape, due to the BAO ratio fit, but the flat DGP model misses some of the uncertainty margins of the data points.  The curved DGP$+k$ model is more visually competitive with $\Lambda$CDM model when fit to BAO in Figure \ref{fig:CosmoEJSk} (middle left), so again it is necessary to rely on the statistical covariance matrix fit to find the preferred model.  Figures \ref{fig:CosmoEJS} (middle right) and \ref{fig:CosmoEJSk} (middle right) show a similar comparison between the models, as expected, since the AP tests are used to calibrate the BAO data.  In Figure \ref{fig:CosmoEJS} (bottom left) and \ref{fig:CosmoEJSk} (bottom left), we see yet another deviation comes from the growth parameter data.  Upon further inspection, we observe the $\Lambda$CDM model may fall outside the error margins of one or two data points, but the DGP models have more difficulty with this data sets that assume a fiducial background as evidenced by the statistical fits from section \ref{UsingCosmoMC}.  

Finally, in Figures \ref{fig:CosmoEJS} (bottom right) and  \ref{fig:CosmoEJSk} (bottom right), not only does the inflection, or transition from deceleration to acceleration of the universe's expansion become evident at different redshift for the different models, but we witness the age of the universe today, and a future crossing or equating of the age of $\Lambda$CDM model with the curved DGP$+k$ model.  Interestingly, other plots like the one in Figure \ref{fig:CosmoEJSk} (bottom right), can show all the GRDE models will having a crossing with a particular DGP model.  In Figure \ref{fig:CosmoEJS} (bottom right), the flat DGP model does not have this same crossing in the future but deviates dramatically from the $\Lambda$CDM model. This final \cosmoejs panel of the expansion versus time allows one to see the different redshift inflections and accelerations of the models in comparison to each other, as well as how the model evolves to the age we observe today.


\section{Conclusion\label{Conclusion}}

In this paper, we showed that the two programs \cosmomc and \cosmoejs can be used to compliment each other in helping to understand the cosmic acceleration.  As an example of the comparable analysis possible with \cosmoejs, we constrained the parameters of cosmological models derived from two popular classes of models, GRDE and DGP, using \cosmomc with recent data sets.  We found comparable results to others in the literature, \cite{PLA2015}, but here, we stress that the values in Table \ref{tb:constraints} can be very useful when recycled into \cosmoejs for simulating cosmological dynamics.  Using \cosmoejs, we can better understand why the models achieved the fits by visually comparing the models to data points.  In particular, we calculated theoretical dynamics for the $\Lambda$CDM, the curved DGP$+k$, and the flat DGP cosmological models, emphasizing which data sets the flat DGP model had difficulty fitting, while simultaneously contrasting it with the more competitive DGP$+k$ model and the $\Lambda$CDM model fits.   Some of these fits are visually similar in the graphing panels of \cosmoejs, which show a need for the statistical covariance matrix fitting methods of Section \ref{UsingCosmoMC} to discern the preferred model.  Not only did the simulations show the graphical and numerical fits (CMB), but we were also able to show the expansion history and future of both models.  Previously, \cosmoejs has been shown to be useful for exploring extreme cases of a particular model. Now, we have shown that we can find best-fit values for cosmological models using \cosmomc and covariance matrix $\chi^2$ methods and then use \cosmoejs to simultaneously compare the dark energy and modified gravity models to each other.  A side-by-side comparison of the models show, not only why those values can be more favorable in comparison to the data, but also show the model's dynamical expansion trajectories.  We will expand the list of provided cosmological models to include more classes of modified gravity models and other varieties of dark energy models.  Also, we will continue to update these simulations with more accurate and precise observations, as they become available.  


\acknowledgments
The authors would like to thank the Donald A. Cowan Physics Institute for support in this document and the development of the \cosmoejs package.  JM would like to thank Larry Engelhardt, Keenan Stone, Zeke Shuler and the Physics and Astronomy department at Francis Marion University for collaboration on earlier versions of the \cosmoejs package.

\bibliographystyle{JHEP}       
\bibliography{ExploringConstraintsCosmoEJSMoldenhauerbibfile}

\begin{thebibliography}{10}
\providecommand{\url}[1]{{#1}}
\providecommand{\urlprefix}{URL }
\expandafter\ifx\csname urlstyle\endcsname\relax
  \providecommand{\doi}[1]{DOI \discretionary{}{}{}#1}\else
  \providecommand{\doi}{DOI \discretionary{}{}{}\begingroup
  \urlstyle{rm}\Url}\fi

\bibitem{SNeDiscovery1}
A.G. Riess, The Astronomical Journal \textbf{116}(3), 1009 (1998).
\newblock \urlprefix\url{http://stacks.iop.org/1538-3881/116/i=3/a=1009}

\bibitem{SNeDiscovery2}
B.P. Schmidt, The Astrophysical Journal \textbf{507}(1), 46 (1998).
\newblock \urlprefix\url{http://stacks.iop.org/0004-637X/507/i=1/a=46}

\bibitem{SNeDiscovery3}
S.~Perlmutter, G.~Aldering, G.~Goldhaber, R.A. Knop, P.~Nugent, P.G. Castro,
  S.~Deustua, S.~Fabbro, A.~Goobar, D.E. Groom, I.M. Hook, A.G. Kim, M.Y. Kim,
  J.C. Lee, N.J. Nunes, R.~Pain, C.R. Pennypacker, R.~Quimby, C.~Lidman, R.S.
  Ellis, M.~Irwin, R.G. McMahon, P.~Ruiz-Lapuente, N.~Walton, B.~Schaefer, B.J.
  Boyle, A.V. Filippenko, T.~Matheson, A.S. Fruchter, N.~Panagia, H.J.M.
  Newberg, W.J. Couch, T.S.C. Project, The Astrophysical Journal
  \textbf{517}(2), 565 (1999).
\newblock \urlprefix\url{http://stacks.iop.org/0004-637X/517/i=2/a=565}

\bibitem{IshakReview2007}
M.~Ishak, Foundations of Physics \textbf{37}(10), 1470 (2007).
\newblock \doi{10.1007/s10701-007-9175-z}.
\newblock \urlprefix\url{http://dx.doi.org/10.1007/s10701-007-9175-z}.
\newblock ID: Ishak2007

\bibitem{PLA2015}
{Planck Collaboration}, {Ade, P. A. R.}, {Aghanim, N.}, {Arnaud, M.}, {Ashdown,
  M.}, {Aumont, J.}, {Baccigalupi, C.}, {Banday, A. J.}, {Barreiro, R. B.},
  {Bartlett, J. G.}, {Bartolo, N.}, {Battaner, E.}, {Battye, R.}, {Benabed,
  K.}, {Benoît, A.}, {Benoit-Lévy, A.}, {Bernard, J.-P.}, {Bersanelli, M.},
  {Bielewicz, P.}, {Bock, J. J.}, {Bonaldi, A.}, {Bonavera, L.}, {Bond, J. R.},
  {Borrill, J.}, {Bouchet, F. R.}, {Boulanger, F.}, {Bucher, M.}, {Burigana,
  C.}, {Butler, R. C.}, {Calabrese, E.}, {Cardoso, J.-F.}, {Catalano, A.},
  {Challinor, A.}, {Chamballu, A.}, {Chary, R.-R.}, {Chiang, H. C.}, {Chluba,
  J.}, {Christensen, P. R.}, {Church, S.}, {Clements, D. L.}, {Colombi, S.},
  {Colombo, L. P. L.}, {Combet, C.}, {Coulais, A.}, {Crill, B. P.}, {Curto,
  A.}, {Cuttaia, F.}, {Danese, L.}, {Davies, R. D.}, {Davis, R. J.}, {de
  Bernardis, P.}, {de Rosa, A.}, {de Zotti, G.}, {Delabrouille, J.}, {Désert,
  F.-X.}, {Di Valentino, E.}, {Dickinson, C.}, {Diego, J. M.}, {Dolag, K.},
  {Dole, H.}, {Donzelli, S.}, {Doré, O.}, {Douspis, M.}, {Ducout, A.},
  {Dunkley, J.}, {Dupac, X.}, {Efstathiou, G.}, {Elsner, F.}, {Enßlin, T. A.},
  {Eriksen, H. K.}, {Farhang, M.}, {Fergusson, J.}, {Finelli, F.}, {Forni, O.},
  {Frailis, M.}, {Fraisse, A. A.}, {Franceschi, E.}, {Frejsel, A.}, {Galeotta,
  S.}, {Galli, S.}, {Ganga, K.}, {Gauthier, C.}, {Gerbino, M.}, {Ghosh, T.},
  {Giard, M.}, {Giraud-Héraud, Y.}, {Giusarma, E.}, {Gjerløw, E.},
  {González-Nuevo, J.}, {Górski, K. M.}, {Gratton, S.}, {Gregorio, A.},
  {Gruppuso, A.}, {Gudmundsson, J. E.}, {Hamann, J.}, {Hansen, F. K.}, {Hanson,
  D.}, {Harrison, D. L.}, {Helou, G.}, {Henrot-Versillé, S.},
  {Hernández-Monteagudo, C.}, {Herranz, D.}, {Hildebrandt, S. R.}, {Hivon, E.},
  {Hobson, M.}, {Holmes, W. A.}, {Hornstrup, A.}, {Hovest, W.}, {Huang, Z.},
  {Huffenberger, K. M.}, {Hurier, G.}, {Jaffe, A. H.}, {Jaffe, T. R.}, {Jones,
  W. C.}, {Juvela, M.}, {Keihänen, E.}, {Keskitalo, R.}, {Kisner, T. S.},
  {Kneissl, R.}, {Knoche, J.}, {Knox, L.}, {Kunz, M.}, {Kurki-Suonio, H.},
  {Lagache, G.}, {Lähteenmäki, A.}, {Lamarre, J.-M.}, {Lasenby, A.}, {Lattanzi,
  M.}, {Lawrence, C. R.}, {Leahy, J. P.}, {Leonardi, R.}, {Lesgourgues, J.},
  {Levrier, F.}, {Lewis, A.}, {Liguori, M.}, {Lilje, P. B.}, {Linden-Vørnle,
  M.}, {López-Caniego, M.}, {Lubin, P. M.}, {Macías-Pérez, J. F.}, {Maggio,
  G.}, {Maino, D.}, {Mandolesi, N.}, {Mangilli, A.}, {Marchini, A.}, {Maris,
  M.}, {Martin, P. G.}, {Martinelli, M.}, {Martínez-González, E.}, {Masi, S.},
  {Matarrese, S.}, {McGehee, P.}, {Meinhold, P. R.}, {Melchiorri, A.}, {Melin,
  J.-B.}, {Mendes, L.}, {Mennella, A.}, {Migliaccio, M.}, {Millea, M.}, {Mitra,
  S.}, {Miville-Deschênes, M.-A.}, {Moneti, A.}, {Montier, L.}, {Morgante, G.},
  {Mortlock, D.}, {Moss, A.}, {Munshi, D.}, {Murphy, J. A.}, {Naselsky, P.},
  {Nati, F.}, {Natoli, P.}, {Netterfield, C. B.}, {Nørgaard-Nielsen, H. U.},
  {Noviello, F.}, {Novikov, D.}, {Novikov, I.}, {Oxborrow, C. A.}, {Paci, F.},
  {Pagano, L.}, {Pajot, F.}, {Paladini, R.}, {Paoletti, D.}, {Partridge, B.},
  {Pasian, F.}, {Patanchon, G.}, {Pearson, T. J.}, {Perdereau, O.}, {Perotto,
  L.}, {Perrotta, F.}, {Pettorino, V.}, {Piacentini, F.}, {Piat, M.},
  {Pierpaoli, E.}, {Pietrobon, D.}, {Plaszczynski, S.}, {Pointecouteau, E.},
  {Polenta, G.}, {Popa, L.}, {Pratt, G. W.}, {Prézeau, G.}, {Prunet, S.},
  {Puget, J.-L.}, {Rachen, J. P.}, {Reach, W. T.}, {Rebolo, R.}, {Reinecke,
  M.}, {Remazeilles, M.}, {Renault, C.}, {Renzi, A.}, {Ristorcelli, I.},
  {Rocha, G.}, {Rosset, C.}, {Rossetti, M.}, {Roudier, G.}, {Rouillé d?Orfeuil,
  B.}, {Rowan-Robinson, M.}, {Rubiño-Martín, J. A.}, {Rusholme, B.}, {Said,
  N.}, {Salvatelli, V.}, {Salvati, L.}, {Sandri, M.}, {Santos, D.},
  {Savelainen, M.}, {Savini, G.}, {Scott, D.}, {Seiffert, M. D.}, {Serra, P.},
  {Shellard, E. P. S.}, {Spencer, L. D.}, {Spinelli, M.}, {Stolyarov, V.},
  {Stompor, R.}, {Sudiwala, R.}, {Sunyaev, R.}, {Sutton, D.}, {Suur-Uski,
  A.-S.}, {Sygnet, J.-F.}, {Tauber, J. A.}, {Terenzi, L.}, {Toffolatti, L.},
  {Tomasi, M.}, {Tristram, M.}, {Trombetti, T.}, {Tucci, M.}, {Tuovinen, J.},
  {Türler, M.}, {Umana, G.}, {Valenziano, L.}, {Valiviita, J.}, {Van Tent, F.},
  {Vielva, P.}, {Villa, F.}, {Wade, L. A.}, {Wandelt, B. D.}, {Wehus, I. K.},
  {White, M.}, {White, S. D. M.}, {Wilkinson, A.}, {Yvon, D.}, {Zacchei, A.},
  {Zonca, A.}, A\&A \textbf{594}, A13 (2016).
\newblock \doi{10.1051/0004-6361/201525830}.
\newblock \urlprefix\url{https://doi.org/10.1051/0004-6361/201525830}

\bibitem{RiessH0}
A.G. Riess, L.M. Macri, S.L. Hoffmann, D.~Scolnic, S.~Casertano, A.V.
  Filippenko, B.E. Tucker, M.J. Reid, D.O. Jones, J.M. Silverman, R.~Chornock,
  P.~Challis, W.~Yuan, P.J. Brown, R.J. Foley, The Astrophysical Journal
  \textbf{826}(1), 56 (2016).
\newblock \urlprefix\url{http://stacks.iop.org/0004-637X/826/i=1/a=56}

\bibitem{CosmoMCPaper}
A.~Lewis, S.~Bridle, Physical Review D \textbf{66}(10), 103511 (2002).
\newblock \urlprefix\url{https://link.aps.org/doi/10.1103/PhysRevD.66.103511}.
\newblock ID: 10.1103/PhysRevD.66.103511; J1: PRD

\bibitem{CosmoNest}
D.~{Parkinson}, P.~{Mukherjee}, A.~{Liddle}.
\newblock {CosmoNest: Cosmological Nested Sampling}.
\newblock Astrophysics Source Code Library (2011)

\bibitem{CosmoPMC}
K.~{Kilbinger}, K.~{Benabed}, O.~{Cappe}, J.~{Cardoso}, J.~{coupon}, G.~{Fort},
  H.J. {McCracken}, S.~{Prunet}, C.P. {Robert}, D.~{Wraith}, arXiv:1101.0950
  pp. 1--64 (2012)

\bibitem{iCosmo}
A.~Refregier, A.~Amara, T.D. Kitching, A.~Rassat, A\&A \textbf{528} (2011).
\newblock \urlprefix\url{https://doi.org/10.1051/0004-6361/200811112}.
\newblock ID: 10.105100046361200811112

\bibitem{CosmoCalcApp}
E.~{Rykoff}, Mobile application software \textbf{Version 2.4} (2012).
\newblock \urlprefix\url{http://itunes.apple.com}

\bibitem{CosmoFish}
M.~Ravelli, M.~Martinelli, arXiv:1606.06268  (2016).
\newblock \urlprefix\url{https://arxiv.org/abs/1606.06268}

\bibitem{EFTCAMB}
B.~Hu, M.~Raveri, N.~Frusciante, A.~Silvestri, Phys. Rev. D \textbf{89}, 103530
  (2014).
\newblock \doi{10.1103/PhysRevD.89.103530}.
\newblock \urlprefix\url{https://link.aps.org/doi/10.1103/PhysRevD.89.103530}

\bibitem{MGCAMB}
G.B. Zhao, L.~Pogosian, A.~Silvestri, J.~Zylberberg, Phys. Rev. D \textbf{79},
  083513 (2009).
\newblock \doi{10.1103/PhysRevD.79.083513}.
\newblock \urlprefix\url{https://link.aps.org/doi/10.1103/PhysRevD.79.083513}

\bibitem{MGCAMB2}
A.~Hojjati, L.~Pogosian, G.B. Zhao, Journal of Cosmology and Astroparticle
  Physics \textbf{2011}(08), 005 (2011).
\newblock \urlprefix\url{http://stacks.iop.org/1475-7516/2011/i=08/a=005}

\bibitem{Rindler}
W.~{Rindler}.
\newblock Relativity: Special, general and cosmological, second edition (2006)

\bibitem{Schneider}
P.~{Schneider}.
\newblock Extragalactic astronomy and cosmology, second edition (2015)

\bibitem{BergstromGoobar}
L.~{Bergstrom}, A.~{Goobar}.
\newblock Cosmology and particle astrophysics, second edition (2004)

\bibitem{Moore}
T.~{Moore}.
\newblock A general relativity workbook (2013)

\bibitem{WrightCalculator}
E.L. Wright, Publications of the Astronomical Society of the Pacific
  \textbf{118}(850), 1711 (2006).
\newblock \urlprefix\url{http://stacks.iop.org/1538-3873/118/i=850/a=1711}

\bibitem{CosmoEJSPaper}
J.~Moldenhauer, L.~Engelhardt, K.M. Stone, E.~Shuler, American Journal of
  Physics \textbf{81}(6), 414 (2013).
\newblock \doi{10.1119/1.4798490}.
\newblock
  \urlprefix\url{https://dbproxy.udallas.edu/login?url=http://search.ebscohost.com/login.aspx?direct=true&db=a9h&AN=88035799&site=ehost-live&scope=site}

\bibitem{CPL}
M.~{Chevallier}, D.~{Polarski}, International Journal of Modern Physics D
  \textbf{10}(02), 213 (2001).
\newblock \doi{10.1142/S0218271801000822}.
\newblock \urlprefix\url{http://dx.doi.org/10.1142/S0218271801000822}.
\newblock Doi: 10.1142/S0218271801000822; 02

\bibitem{DGPghostmode}
K.~Koyama, Phys. Rev. D \textbf{72},
  123511 (2005).
\newblock \doi{10.1103/PhysRevD.72.123511}.
\newblock \urlprefix\url{https://link.aps.org/doi/10.1103/PhysRevD.72.123511}

\bibitem{Fang2008}
W.~Fang, S.~Wang, W.~Hu, Z.~Haiman, L.~Ham, M.~May, Phys. Rev. D \textbf{78},
  103509 (2008).
\newblock \doi{10.1103/PhysRevD.78.103509}.
\newblock \urlprefix\url{https://link.aps.org/doi/10.1103/PhysRevD.78.103509}

\bibitem{Dossett2010}
J.~Dossett, M.~Ishak, J.~Moldenhauer, A.W. Yungui Gong~and, Journal of
  Cosmology and Astroparticle Physics \textbf{2010}(04), 022 (2010).
\newblock \urlprefix\url{http://stacks.iop.org/1475-7516/2010/i=04/a=022}

\bibitem{DGP}
G.~Dvali, G.~Gabadadze, M.~Porrati, Physics Letters B \textbf{485}(1), 208
  (2000).
\newblock \doi{http://dx.doi.org/10.1016/S0370-2693(00)00669-9}.
\newblock
  \urlprefix\url{http://www.sciencedirect.com/science/article/pii/S0370269300006699}

\bibitem{Union2.1}
Yee, N.~The Supernova Cosmology~Project, D.Rubin, C.Lidman, G.Aldering,
  R.Amanullah, K.Barbary, L.F.Barrientos, J.Botyanszki, M.Brodwin, N.Connolly,
  K.S.Dawson, A.Dey, M.Doi, M.Donahue, S.Deustua, P.Eisenhardt, E.Ellingson,
  L.Faccioli, V.Fadeyev, H.K.Fakhouri, A.S.Fruchter, D.G.Gilbank, M.D.Gladders,
  G.Goldhaber, A.H.Gonzalez, A.Goobar, A.Gude, T.Hattori, H.Hoekstra, E.Hsiao,
  X.Huang, Y.Ihara, M.J.Jee, D.Johnston, N.Kashikawa, B.Koester, K.Konishi,
  M.Kowalski, E.V.Linder, L.Lubin, J.Melbourne, J.Meyers, T.Morokuma, F.Munshi,
  C.Mullis, T.Oda, N.Panagia, S.Perlmutter, M.Postman, T.Pritchard, J.Rhodes,
  P.Ripoche, P.Rosati, D.J.Schlegel, A.Spadafora, S.A.Stanford, V.Stanishev,
  D.Stern, M.Strovink, N.Takanashi, K.Tokita, M.Wagner, L.Wang, N.Yasuda,
  H.K.C., The Astrophysical Journal \textbf{746}(1), 85 (2012).
\newblock \urlprefix\url{http://stacks.iop.org/0004-637X/746/i=1/a=85}

\bibitem{JLA}
{Betoule, M.}, {Kessler, R.}, {Guy, J.}, {Mosher, J.}, {Hardin, D.}, {Biswas,
  R.}, {Astier, P.}, {El-Hage, P.}, {Konig, M.}, {Kuhlmann, S.}, {Marriner,
  J.}, {Pain, R.}, {Regnault, N.}, {Balland, C.}, {Bassett, B. A.}, {Brown, P.
  J.}, {Campbell, H.}, {Carlberg, R. G.}, {Cellier-Holzem, F.}, {Cinabro, D.},
  {Conley, A.}, {D?Andrea, C. B.}, {DePoy, D. L.}, {Doi, M.}, {Ellis, R. S.},
  {Fabbro, S.}, {Filippenko, A. V.}, {Foley, R. J.}, {Frieman, J. A.},
  {Fouchez, D.}, {Galbany, L.}, {Goobar, A.}, {Gupta, R. R.}, {Hill, G. J.},
  {Hlozek, R.}, {Hogan, C. J.}, {Hook, I. M.}, {Howell, D. A.}, {Jha, S. W.},
  {Le Guillou, L.}, {Leloudas, G.}, {Lidman, C.}, {Marshall, J. L.}, {Möller,
  A.}, {Mourão, A. M.}, {Neveu, J.}, {Nichol, R.}, {Olmstead, M. D.},
  {Palanque-Delabrouille, N.}, {Perlmutter, S.}, {Prieto, J. L.}, {Pritchet, C.
  J.}, {Richmond, M.}, {Riess, A. G.}, {Ruhlmann-Kleider, V.}, {Sako, M.},
  {Schahmaneche, K.}, {Schneider, D. P.}, {Smith, M.}, {Sollerman, J.},
  {Sullivan, M.}, {Walton, N. A.}, {Wheeler, C. J.}, A\&A \textbf{568}, A22
  (2014).
\newblock \doi{10.1051/0004-6361/201423413}.
\newblock \urlprefix\url{https://doi.org/10.1051/0004-6361/201423413}

\bibitem{LSST}
L.S.S.T. {Collaboration}.
\newblock Large synoptic survey telescope (lsst).
\newblock \urlprefix\url{http://www.lsst.org}

\bibitem{HzCCvBAO}
K.~{Leaf}, F.~{Melia}, \mnras pp. 1--17 (2017)

\bibitem{Zhangs}
C.~Zhang, H.~Zhang, S.~Yuan, S.~Liu, a.Y. Tong-Jie~Zhang, Research in Astronomy
  and Astrophysics \textbf{14}(10), 1221 (2014).
\newblock \urlprefix\url{http://stacks.iop.org/1674-4527/14/i=10/a=002}

\bibitem{MaStro}
M.~Moresco, L.~Verde, L.~Pozzetti, A.C. Raul Jimenez~and, Journal of Cosmology
  and Astroparticle Physics \textbf{2012}(07), 053 (2012).
\newblock \urlprefix\url{http://stacks.iop.org/1475-7516/2012/i=07/a=053}

\bibitem{Ages}
D.~Stern, R.~Jimenez, L.~Verde, S.S. Marc Kamionkowski~and, Journal of
  Cosmology and Astroparticle Physics \textbf{2010}(02), 008 (2010).
\newblock \urlprefix\url{http://stacks.iop.org/1475-7516/2010/i=02/a=008}

\bibitem{BlakeHz}
C.~Blake, S.~Brough, M.~Colless, C.~Contreras, W.~Couch, S.~Croom, D.~Croton,
  T.M. Davis, M.J. Drinkwater, K.~Forster, D.~Gilbank, M.~Gladders,
  K.~Glazebrook, B.~Jelliffe, R.J. Jurek, I.h. Li, B.~Madore, D.C. Martin,
  K.~Pimbblet, G.B. Poole, M.~Pracy, R.~Sharp, E.~Wisnioski, D.~Woods, T.K.
  Wyder, H.K.C. Yee, Monthly Notices of the Royal Astronomical Society
  \textbf{425}(1), 405 (2012).
\newblock \doi{10.1111/j.1365-2966.2012.21473.x}.
\newblock \urlprefix\url{+ http://dx.doi.org/10.1111/j.1365-2966.2012.21473.x}

\bibitem{SDSSHz}
E.~Gaztañaga, A.~Cabré, L.~Hui, Monthly Notices of the Royal Astronomical
  Society \textbf{399}(3), 1663 (2009).
\newblock \urlprefix\url{http://dx.doi.org/10.1111/j.1365-2966.2009.15405.x}.
\newblock 10.1111/j.1365-2966.2009.15405.x

\bibitem{OkaHz}
A.~Oka, S.~Saito, T.~Nishimichi, A.~Taruya, K.~Yamamoto, Monthly Notices of the
  Royal Astronomical Society \textbf{439}(3), 2515 (2014).
\newblock \urlprefix\url{http://dx.doi.org/10.1093/mnras/stu111}.
\newblock 10.1093/mnras/stu111

\bibitem{M2015}
M.~Moresco, Monthly Notices of the Royal Astronomical Society: Letters
  \textbf{450}(1), L16 (2015).
\newblock \doi{10.1093/mnrasl/slv037}.
\newblock \urlprefix\url{+ http://dx.doi.org/10.1093/mnrasl/slv037}

\bibitem{ChuangHzSDSSR7}
C.H. Chuang, Y.~Wang, Monthly Notices of the Royal Astronomical Society
  \textbf{435}(1), 255 (2013).
\newblock \urlprefix\url{http://dx.doi.org/10.1093/mnras/stt1290}.
\newblock 10.1093/mnras/stt1290

\bibitem{BDR}
{Busca, N. G.}, {Delubac, T.}, {Rich, J.}, {Bailey, S.}, {Font-Ribera, A.},
  {Kirkby, D.}, {Le Goff, J.-M.}, {Pieri, M. M.}, {Slosar, A.}, {Aubourg, É.},
  {Bautista, J. E.}, {Bizyaev, D.}, {Blomqvist, M.}, {Bolton, A. S.}, {Bovy,
  J.}, {Brewington, H.}, {Borde, A.}, {Brinkmann, J.}, {Carithers, B.}, {Croft,
  R. A. C.}, {Dawson, K. S.}, {Ebelke, G.}, {Eisenstein, D. J.}, {Hamilton,
  J.-C.}, {Ho, S.}, {Hogg, D. W.}, {Honscheid, K.}, {Lee, K.-G.}, {Lundgren,
  B.}, {Malanushenko, E.}, {Malanushenko, V.}, {Margala, D.}, {Maraston, C.},
  {Mehta, K.}, {Miralda-Escudé, J.}, {Myers, A. D.}, {Nichol, R. C.},
  {Noterdaeme, P.}, {Olmstead, M. D.}, {Oravetz, D.}, {Palanque-Delabrouille,
  N.}, {Pan, K.}, {Pâris, I.}, {Percival, W. J.}, {Petitjean, P.}, {Roe, N.
  A.}, {Rollinde, E.}, {Ross, N. P.}, {Rossi, G.}, {Schlegel, D. J.},
  {Schneider, D. P.}, {Shelden, A.}, {Sheldon, E. S.}, {Simmons, A.}, {Snedden,
  S.}, {Tinker, J. L.}, {Viel, M.}, {Weaver, B. A.}, {Weinberg, D. H.}, {White,
  M.}, {Yèche, C.}, {York, D. G.}, A\&A \textbf{552}, A96 (2013).
\newblock \doi{10.1051/0004-6361/201220724}.
\newblock \urlprefix\url{https://doi.org/10.1051/0004-6361/201220724}

\bibitem{DelubacHz}
{Delubac, Timothée}, {Bautista, Julian E.}, {Busca, Nicolás G.}, {Rich, James},
  {Kirkby, David}, {Bailey, Stephen}, {Font-Ribera, Andreu}, {Slosar, An?e},
  {Lee, Khee-Gan}, {Pieri, Matthew M.}, {Hamilton, Jean-Christophe}, {Aubourg,
  Éric}, {Blomqvist, Michael}, {Bovy, Jo}, {Brinkmann, Jon}, {Carithers,
  William}, {Dawson, Kyle S.}, {Eisenstein, Daniel J.}, {Gontcho A Gontcho,
  Satya}, {Kneib, Jean-Paul}, {Le Goff, Jean-Marc}, {Margala, Daniel},
  {Miralda-Escudé, Jordi}, {Myers, Adam D.}, {Nichol, Robert C.}, {Noterdaeme,
  Pasquier}, {O?Connell, Ross}, {Olmstead, Matthew D.}, {Palanque-Delabrouille,
  Nathalie}, {Pâris, Isabelle}, {Petitjean, Patrick}, {Ross, Nicholas P.},
  {Rossi, Graziano}, {Schlegel, David J.}, {Schneider, Donald P.}, {Weinberg,
  David H.}, {Yèche, Christophe}, {York, Donald G.}, A\&A \textbf{574}, A59
  (2015).
\newblock \doi{10.1051/0004-6361/201423969}.
\newblock \urlprefix\url{https://doi.org/10.1051/0004-6361/201423969}

\bibitem{AndersonHzBOSS}
L.~Anderson, Ã.~Aubourg, S.~Bailey, F.~Beutler, V.~Bhardwaj, M.~Blanton, A.S.
  Bolton, J.~Brinkmann, J.R. Brownstein, A.~Burden, C.H. Chuang, A.J. Cuesta,
  K.S. Dawson, D.J. Eisenstein, S.~Escoffier, J.E. Gunn, H.~Guo, S.~Ho,
  K.~Honscheid, C.~Howlett, D.~Kirkby, R.H. Lupton, M.~Manera, C.~Maraston,
  C.K. McBride, O.~Mena, F.~Montesano, R.C. Nichol, S.E. Nuza, M.D. Olmstead,
  N.~Padmanabhan, N.~Palanque-Delabrouille, J.~Parejko, W.J. Percival,
  P.~Petitjean, F.~Prada, A.~Price-Whelan, B.~Reid, N.A. Roe, A.J. Ross, N.P.
  Ross, C.G. Sabiu, S.~Saito, L.~Samushia, A.G. Sánchez, D.J. Schlegel, D.P.
  Schneider, C.G. Scoccola, H.J. Seo, R.A. Skibba, M.A. Strauss, M.E.C.
  Swanson, D.~Thomas, J.L. Tinker, R.~Tojeiro, M.V. Magaña, L.~Verde, D.A.
  Wake, B.A. Weaver, D.H. Weinberg, M.~White, X.~Xu, C.~YÃ?che, I.~Zehavi, G.B.
  Zhao, Monthly Notices of the Royal Astronomical Society \textbf{441}(1), 24
  (2014).
\newblock \urlprefix\url{http://dx.doi.org/10.1093/mnras/stu523}.
\newblock 10.1093/mnras/stu523

\bibitem{FontRiberaBOSSDR11}
A.~Font-Ribera, D.~Kirkby, N.~Busca, J.~Miralda-Escudé, N.P. Ross, A.~Slosar,
  J.~Rich, Ã?ric Aubourg, S.~Bailey, V.~Bhardwaj, J.~Bautista, F.~Beutler,
  D.~Bizyaev, M.~Blomqvist, H.~Brewington, J.~Brinkmann, J.R. Brownstein,
  B.~Carithers, K.S. Dawson, T.~Delubac, G.~Ebelke, D.J. Eisenstein, J.~Ge,
  K.~Kinemuchi, K.G. Lee, V.~Malanushenko, E.~Malanushenko, M.~Marchante,
  D.~Margala, D.~Muna, A.D. Myers, P.~Noterdaeme, D.~Oravetz,
  N.~Palanque-Delabrouille, I.~Pâris, P.~Petitjean, M.M. Pieri, G.~Rossi, D.P.
  Schneider, A.~Simmons, M.~Viel, D.G. Christophe Yeche~and, Journal of
  Cosmology and Astroparticle Physics \textbf{2014}(05), 027 (2014).
\newblock \urlprefix\url{http://stacks.iop.org/1475-7516/2014/i=05/a=027}

\bibitem{PercivalBAO}
W.J. Percival, B.A. Reid, D.J. Eisenstein, N.A. Bahcall, T.~Budavari, J.A.
  Frieman, M.~Fukugita, J.E. Gunn, ??eljko Ivezi??, G.R. Knapp, R.G. Kron,
  J.~Loveday, R.H. Lupton, T.A. McKay, A.~Meiksin, R.C. Nichol, A.C. Pope, D.J.
  Schlegel, D.P. Schneider, D.N. Spergel, C.~Stoughton, M.A. Strauss, A.S.
  Szalay, M.~Tegmark, M.S. Vogeley, D.H. Weinberg, D.G. York, I.~Zehavi,
  Monthly Notices of the Royal Astronomical Society \textbf{401}(4), 2148
  (2010).
\newblock \urlprefix\url{http://dx.doi.org/10.1111/j.1365-2966.2009.15812.x}.
\newblock 10.1111/j.1365-2966.2009.15812.x

\bibitem{WiggleZBAO}
C.~Blake, T.~Davis, G.B. Poole, D.~Parkinson, S.~Brough, M.~Colless,
  C.~Contreras, W.~Couch, S.~Croom, M.J. Drinkwater, K.~Forster, D.~Gilbank,
  M.~Gladders, K.~Glazebrook, B.~Jelliffe, R.J. Jurek, I.~hui Li, B.~Madore,
  D.C. Martin, K.~Pimbblet, M.~Pracy, R.~Sharp, E.~Wisnioski, D.~Woods, T.K.
  Wyder, H.K.C. Yee, Monthly Notices of the Royal Astronomical Society
  \textbf{415}(3), 2892 (2011).
\newblock \urlprefix\url{http://dx.doi.org/10.1111/j.1365-2966.2011.19077.x}.
\newblock 10.1111/j.1365-2966.2011.19077.x

\bibitem{Beutler6dFGSBAO}
F.~Beutler, C.~Blake, M.~Colless, D.H. Jones, L.~Staveley-Smith, L.~Campbell,
  Q.~Parker, W.~Saunders, F.~Watson, Monthly Notices of the Royal Astronomical
  Society \textbf{416}(4), 3017 (2011).
\newblock \urlprefix\url{http://dx.doi.org/10.1111/j.1365-2966.2011.19250.x}.
\newblock 10.1111/j.1365-2966.2011.19250.x

\bibitem{Percival2007}
W.J. Percival, S.~Cole, D.J. Eisenstein, R.C. Nichol, J.A. Peacock, A.C. Pope,
  A.S. Szalay, Monthly Notices of the Royal Astronomical Society
  \textbf{381}(3), 1053 (2007).
\newblock \doi{10.1111/j.1365-2966.2007.12268.x}.
\newblock \urlprefix\url{+ http://dx.doi.org/10.1111/j.1365-2966.2007.12268.x}

\bibitem{AndersonBAOBOSS}
L.~Anderson, E.~Aubourg, S.~Bailey, D.~Bizyaev, M.~Blanton, A.S. Bolton,
  J.~Brinkmann, J.R. Brownstein, A.~Burden, A.J. Cuesta, L.A.N. da~Costa, K.S.
  Dawson, R.~de~Putter, D.J. Eisenstein, J.E. Gunn, H.~Guo, J.C. Hamilton,
  P.~Harding, S.~Ho, K.~Honscheid, E.~Kazin, D.~Kirkby, J.P. Kneib, A.~Labatie,
  C.~Loomis, R.H. Lupton, E.~Malanushenko, V.~Malanushenko, R.~Mandelbaum,
  M.~Manera, C.~Maraston, C.K. McBride, K.T. Mehta, O.~Mena, F.~Montesano,
  D.~Muna, R.C. Nichol, S.E. Nuza, M.D. Olmstead, D.~Oravetz, N.~Padmanabhan,
  N.~Palanque-Delabrouille, K.~Pan, J.~Parejko, I.~Pâris, W.J. Percival,
  P.~Petitjean, F.~Prada, B.~Reid, N.A. Roe, A.J. Ross, N.P. Ross, L.~Samushia,
  A.G. Sánchez, D.J. Schlegel, D.P. Schneider, C.G. Scóccola, H.J. Seo, E.S.
  Sheldon, A.~Simmons, R.A. Skibba, M.A. Strauss, M.E.C. Swanson, D.~Thomas,
  J.L. Tinker, R.~Tojeiro, M.V. Magaña, L.~Verde, C.~Wagner, D.A. Wake, B.A.
  Weaver, D.H. Weinberg, M.~White, X.~Xu, C.~Y??che, I.~Zehavi, G.B. Zhao,
  Monthly Notices of the Royal Astronomical Society \textbf{427}(4), 3435
  (2012).
\newblock \urlprefix\url{http://dx.doi.org/10.1111/j.1365-2966.2012.22066.x}.
\newblock 10.1111/j.1365-2966.2012.22066.x

\bibitem{EisensteinHu1998}
D.J. Eisenstein, W.~Hu, The Astrophysical Journal \textbf{496}(2), 605 (1998).
\newblock \urlprefix\url{http://stacks.iop.org/0004-637X/496/i=2/a=605}

\bibitem{Eisenstein2005}
D.J. Eisenstein, I.~Zehavi, D.W. Hogg, R.~Scoccimarro, M.R. Blanton, R.C.
  Nichol, R.~Scranton, H.J. Seo, M.~Tegmark, Z.~Zheng, S.F. Anderson, J.~Annis,
  N.~Bahcall, J.~Brinkmann, S.~Burles, F.J. Castander, A.~Connolly, I.~Csabai,
  M.~Doi, M.~Fukugita, J.A. Frieman, K.~Glazebrook, J.E. Gunn, J.S. Hendry,
  G.~Hennessy, Z.~Ivezi?, S.~Kent, G.R. Knapp, H.~Lin, Y.S. Loh, R.H. Lupton,
  B.~Margon, T.A. McKay, A.~Meiksin, J.A. Munn, A.~Pope, M.W. Richmond,
  D.~Schlegel, D.P. Schneider, K.~Shimasaku, C.~Stoughton, M.A. Strauss,
  M.~SubbaRao, A.S. Szalay, I.~Szapudi, D.L. Tucker, B.~Yanny, D.G. York, The
  Astrophysical Journal \textbf{633}(2), 560 (2005).
\newblock \urlprefix\url{http://stacks.iop.org/0004-637X/633/i=2/a=560}

\bibitem{WangMukherjee}
Y.~Wang, P.~Mukherjee, Phys. Rev. D \textbf{76}, 103533 (2007).
\newblock \doi{10.1103/PhysRevD.76.103533}.
\newblock \urlprefix\url{https://link.aps.org/doi/10.1103/PhysRevD.76.103533}

\bibitem{Wright}
E.L. Wright, The Astrophysical Journal \textbf{664}(2), 633 (2007).
\newblock \urlprefix\url{http://stacks.iop.org/0004-637X/664/i=2/a=633}

\bibitem{HuSugiyama}
W.~Hu, N.~Sugiyama, The Astrophysical Journal \textbf{471}(2), 542 (1996).
\newblock \urlprefix\url{http://stacks.iop.org/0004-637X/471/i=2/a=542}

\bibitem{Bond1997}
J.R. {Bond}, G.~{Efstathiou}, M.~{Tegmark}, \mnras \textbf{291}, L33 (1997).
\newblock \doi{10.1093/mnras/291.1.L33}

\bibitem{Wang2013}
Y.~Wang, S.~Wang, Phys. Rev. D \textbf{88}, 043522 (2013).
\newblock \doi{10.1103/PhysRevD.88.043522}.
\newblock \urlprefix\url{https://link.aps.org/doi/10.1103/PhysRevD.88.043522}

\bibitem{PLA2013}
{Planck Collaboration}, {Ade, P. A. R.}, {Aghanim, N.}, {Armitage-Caplan, C.},
  {Arnaud, M.}, {Ashdown, M.}, {Atrio-Barandela, F.}, {Aumont, J.},
  {Baccigalupi, C.}, {Banday, A. J.}, {Barreiro, R. B.}, {Bartlett, J. G.},
  {Battaner, E.}, {Benabed, K.}, {Benoît, A.}, {Benoit-Lévy, A.}, {Bernard,
  J.-P.}, {Bersanelli, M.}, {Bielewicz, P.}, {Bobin, J.}, {Bock, J. J.},
  {Bonaldi, A.}, {Bond, J. R.}, {Borrill, J.}, {Bouchet, F. R.}, {Bridges, M.},
  {Bucher, M.}, {Burigana, C.}, {Butler, R. C.}, {Calabrese, E.}, {Cappellini,
  B.}, {Cardoso, J.-F.}, {Catalano, A.}, {Challinor, A.}, {Chamballu, A.},
  {Chary, R.-R.}, {Chen, X.}, {Chiang, H. C.}, {Chiang, L.-Y}, {Christensen, P.
  R.}, {Church, S.}, {Clements, D. L.}, {Colombi, S.}, {Colombo, L. P. L.},
  {Couchot, F.}, {Coulais, A.}, {Crill, B. P.}, {Curto, A.}, {Cuttaia, F.},
  {Danese, L.}, {Davies, R. D.}, {Davis, R. J.}, {de Bernardis, P.}, {de Rosa,
  A.}, {de Zotti, G.}, {Delabrouille, J.}, {Delouis, J.-M.}, {Désert, F.-X.},
  {Dickinson, C.}, {Diego, J. M.}, {Dolag, K.}, {Dole, H.}, {Donzelli, S.},
  {Doré, O.}, {Douspis, M.}, {Dunkley, J.}, {Dupac, X.}, {Efstathiou, G.},
  {Elsner, F.}, {Enßlin, T. A.}, {Eriksen, H. K.}, {Finelli, F.}, {Forni, O.},
  {Frailis, M.}, {Fraisse, A. A.}, {Franceschi, E.}, {Gaier, T. C.}, {Galeotta,
  S.}, {Galli, S.}, {Ganga, K.}, {Giard, M.}, {Giardino, G.}, {Giraud-Héraud,
  Y.}, {Gjerløw, E.}, {González-Nuevo, J.}, {Górski, K. M.}, {Gratton, S.},
  {Gregorio, A.}, {Gruppuso, A.}, {Gudmundsson, J. E.}, {Haissinski, J.},
  {Hamann, J.}, {Hansen, F. K.}, {Hanson, D.}, {Harrison, D.},
  {Henrot-Versillé, S.}, {Hernández-Monteagudo, C.}, {Herranz, D.},
  {Hildebrandt, S. R.}, {Hivon, E.}, {Hobson, M.}, {Holmes, W. A.}, {Hornstrup,
  A.}, {Hou, Z.}, {Hovest, W.}, {Huffenberger, K. M.}, {Jaffe, A. H.}, {Jaffe,
  T. R.}, {Jewell, J.}, {Jones, W. C.}, {Juvela, M.}, {Keihänen, E.},
  {Keskitalo, R.}, {Kisner, T. S.}, {Kneissl, R.}, {Knoche, J.}, {Knox, L.},
  {Kunz, M.}, {Kurki-Suonio, H.}, {Lagache, G.}, {Lähteenmäki, A.}, {Lamarre,
  J.-M.}, {Lasenby, A.}, {Lattanzi, M.}, {Laureijs, R. J.}, {Lawrence, C. R.},
  {Leach, S.}, {Leahy, J. P.}, {Leonardi, R.}, {León-Tavares, J.},
  {Lesgourgues, J.}, {Lewis, A.}, {Liguori, M.}, {Lilje, P. B.},
  {Linden-Vørnle, M.}, {López-Caniego, M.}, {Lubin, P. M.}, {Macías-Pérez, J.
  F.}, {Maffei, B.}, {Maino, D.}, {Mandolesi, N.}, {Maris, M.}, {Marshall, D.
  J.}, {Martin, P. G.}, {Martínez-González, E.}, {Masi, S.}, {Massardi, M.},
  {Matarrese, S.}, {Matthai, F.}, {Mazzotta, P.}, {Meinhold, P. R.},
  {Melchiorri, A.}, {Melin, J.-B.}, {Mendes, L.}, {Menegoni, E.}, {Mennella,
  A.}, {Migliaccio, M.}, {Millea, M.}, {Mitra, S.}, {Miville-Deschênes, M.-A.},
  {Moneti, A.}, {Montier, L.}, {Morgante, G.}, {Mortlock, D.}, {Moss, A.},
  {Munshi, D.}, {Murphy, J. A.}, {Naselsky, P.}, {Nati, F.}, {Natoli, P.},
  {Netterfield, C. B.}, {Nørgaard-Nielsen, H. U.}, {Noviello, F.}, {Novikov,
  D.}, {Novikov, I.}, {O?Dwyer, I. J.}, {Osborne, S.}, {Oxborrow, C. A.},
  {Paci, F.}, {Pagano, L.}, {Pajot, F.}, {Paladini, R.}, {Paoletti, D.},
  {Partridge, B.}, {Pasian, F.}, {Patanchon, G.}, {Pearson, D.}, {Pearson, T.
  J.}, {Peiris, H. V.}, {Perdereau, O.}, {Perotto, L.}, {Perrotta, F.},
  {Pettorino, V.}, {Piacentini, F.}, {Piat, M.}, {Pierpaoli, E.}, {Pietrobon,
  D.}, {Plaszczynski, S.}, {Platania, P.}, {Pointecouteau, E.}, {Polenta, G.},
  {Ponthieu, N.}, {Popa, L.}, {Poutanen, T.}, {Pratt, G. W.}, {Prézeau, G.},
  {Prunet, S.}, {Puget, J.-L.}, {Rachen, J. P.}, {Reach, W. T.}, {Rebolo, R.},
  {Reinecke, M.}, {Remazeilles, M.}, {Renault, C.}, {Ricciardi, S.}, {Riller,
  T.}, {Ristorcelli, I.}, {Rocha, G.}, {Rosset, C.}, {Roudier, G.},
  {Rowan-Robinson, M.}, {Rubiño-Martín, J. A.}, {Rusholme, B.}, {Sandri, M.},
  {Santos, D.}, {Savelainen, M.}, {Savini, G.}, {Scott, D.}, {Seiffert, M. D.},
  {Shellard, E. P. S.}, {Spencer, L. D.}, {Starck, J.-L.}, {Stolyarov, V.},
  {Stompor, R.}, {Sudiwala, R.}, {Sunyaev, R.}, {Sureau, F.}, {Sutton, D.},
  {Suur-Uski, A.-S.}, {Sygnet, J.-F.}, {Tauber, J. A.}, {Tavagnacco, D.},
  {Terenzi, L.}, {Toffolatti, L.}, {Tomasi, M.}, {Tristram, M.}, {Tucci, M.},
  {Tuovinen, J.}, {Türler, M.}, {Umana, G.}, {Valenziano, L.}, {Valiviita, J.},
  {Van Tent, B.}, {Vielva, P.}, {Villa, F.}, {Vittorio, N.}, {Wade, L. A.},
  {Wandelt, B. D.}, {Wehus, I. K.}, {White, M.}, {White, S. D. M.}, {Wilkinson,
  A.}, {Yvon, D.}, {Zacchei, A.}, {Zonca, A.}, A\&A \textbf{571}, A16 (2014).
\newblock \doi{10.1051/0004-6361/201321591}.
\newblock \urlprefix\url{https://doi.org/10.1051/0004-6361/201321591}

\bibitem{IshakDEDGP2006}
M.~Ishak, A.~Upadhye, D.N. Spergel, Physical Review D \textbf{74}(4), 043513
  (2006).
\newblock \urlprefix\url{https://link.aps.org/doi/10.1103/PhysRevD.74.043513}.
\newblock ID: 10.1103/PhysRevD.74.043513; J1: PRD

\bibitem{HuangWang2015}
Q.G. Huang, K.~Wang, S.~Wang, Journal of Cosmology and Astroparticle Physics
  \textbf{2015}(12), 022 (2015).
\newblock \urlprefix\url{http://stacks.iop.org/1475-7516/2015/i=12/a=022}

\bibitem{APOriginal1979}
C.~{Alcock}, B.~{Paczynski}, Nature \textbf{281}, 358 (1979).
\newblock \doi{10.1038/281358a0}

\bibitem{APtestsSamuisha}
L.~Samushia, B.A. Reid, M.~White, W.J. Percival, A.J. Cuesta, G.B. Zhao, A.J.
  Ross, M.~Manera, .~Aubourg, F.~Beutler, J.~Brinkmann, J.R. Brownstein, K.S.
  Dawson, D.J. Eisenstein, S.~Ho, K.~Honscheid, C.~Maraston, F.~Montesano, R.C.
  Nichol, N.A. Roe, N.P. Ross, A.G. Sánchez, D.J. Schlegel, D.P. Schneider,
  A.~Streblyanska, D.~Thomas, J.L. Tinker, D.A. Wake, B.A. Weaver, I.~Zehavi,
  Monthly Notices of the Royal Astronomical Society \textbf{439}(4), 3504
  (2014).
\newblock \urlprefix\url{http://dx.doi.org/10.1093/mnras/stu197}.
\newblock 10.1093/mnras/stu197

\bibitem{SongPercivalRSD}
Y.S. Song, W.J. Percival, Journal of Cosmology and Astroparticle Physics
  \textbf{2009}(10), 004 (2009).
\newblock \urlprefix\url{http://stacks.iop.org/1475-7516/2009/i=10/a=004}

\bibitem{H070p6}
G.~Efstathiou, Monthly Notices of the Royal Astronomical Society
  \textbf{440}(2), 1138 (2014).
\newblock \doi{10.1093/mnras/stu278}.
\newblock \urlprefix\url{http://dx.doi.org/10.1093/mnras/stu278}

\bibitem{Beutler6dFGS}
F.~Beutler, C.~Blake, M.~Colless, D.H. Jones, L.~Staveley-Smith, L.~Campbell,
  Q.~Parker, W.~Saunders, F.~Watson, Monthly Notices of the Royal Astronomical
  Society \textbf{416}(4), 3017 (2011).
\newblock \urlprefix\url{http://dx.doi.org/10.1111/j.1365-2966.2011.19250.x}.
\newblock 10.1111/j.1365-2966.2011.19250.x

\bibitem{RossMGS}
A.J. Ross, L.~Samushia, C.~Howlett, W.J. Percival, A.~Burden, M.~Manera,
  Monthly Notices of the Royal Astronomical Society \textbf{449}(1), 835
  (2015).
\newblock \doi{10.1093/mnras/stv154}.
\newblock \urlprefix\url{+ http://dx.doi.org/10.1093/mnras/stv154}

\bibitem{DR12BOSSCMASS}
H.~Gil-Marín, W.J. Percival, A.J. Cuesta, J.R. Brownstein, C.H. Chuang, S.~Ho,
  F.S. Kitaura, C.~Maraston, F.~Prada, S.~Rodríguez-Torres, A.J. Ross, D.J.
  Schlegel, D.P. Schneider, D.~Thomas, J.L. Tinker, R.~Tojeiro,
  M.~Vargas Magaña, G.B. Zhao, Monthly Notices of the Royal Astronomical
  Society \textbf{460}(4), 4210 (2016).
\newblock \doi{10.1093/mnras/stw1264}.
\newblock \urlprefix\url{+ http://dx.doi.org/10.1093/mnras/stw1264}

\bibitem{DR12BOSSLOWZ}
A.J. Cuesta, M.~Vargas-Magaña, F.~Beutler, A.S. Bolton, J.R. Brownstein, D.J.
  Eisenstein, H.~Gil-Marín, S.~Ho, C.K. McBride, C.~Maraston, N.~Padmanabhan,
  W.J. Percival, B.A. Reid, A.J. Ross, N.P. Ross, A.G. Sánchez, D.J. Schlegel,
  D.P. Schneider, D.~Thomas, J.~Tinker, R.~Tojeiro, L.~Verde, M.~White, Monthly
  Notices of the Royal Astronomical Society \textbf{457}(2), 1770 (2016).
\newblock \doi{10.1093/mnras/stw066}.
\newblock \urlprefix\url{+ http://dx.doi.org/10.1093/mnras/stw066}

\bibitem{H0Riess2016}
A.G. Riess, L.M. Macri, S.L. Hoffmann, D.~Scolnic, S.~Casertano, A.V.
  Filippenko, B.E. Tucker, M.J. Reid, D.O. Jones, J.M. Silverman, R.~Chornock,
  P.~Challis, W.~Yuan, P.J. Brown, R.J. Foley, The Astrophysical Journal
  \textbf{826}(1), 56 (2016).
\newblock \urlprefix\url{http://stacks.iop.org/0004-637X/826/i=1/a=56}

\bibitem{EJS}
W.~Christian, F.~Esquembre, The Physics Teacher \textbf{45}(8), 475 (2007).
\newblock \doi{10.1119/1.2798358}.
\newblock \urlprefix\url{http://dx.doi.org/10.1119/1.2798358}

\end{thebibliography}

\end{document}